\documentclass[12pt,preprint]{aastex}

\usepackage{color}
\usepackage{gensymb}
\usepackage{natbib}
\usepackage{url}
\newcommand{\RNum}[1]{\uppercase\expandafter{\romannumeral #1\relax}}

\slugcomment{in prep.}

\shorttitle{An optical transmission spectrum of GJ~1214b}
\shortauthors{Rackham et al.}

\begin{document}


\title{\textcolor{black}{\textit{ACCESS} I: An Optical Transmission Spectrum of GJ~1214b Reveals a Heterogeneous Stellar Photosphere}}




\author{Benjamin Rackham\altaffilmark{1,2}}
\affil{Department of Astronomy/Steward Observatory, The University of Arizona, 933 N. Cherry Avenue, Tucson, AZ 85721, USA}
\email{brackham@as.arizona.edu}

\author{N\'estor Espinoza}
\affil{Instituto de Astrof\'isica, Facultad de F\'isica, Pontificia Universidad Cat\'olica de Chile, Av.\ Vicu\~na Mackenna 4860, 7820436 Macul, Santiago, Chile}
\affil{Millennium Institute of Astrophysics, Chile}

\author{D\'aniel Apai\altaffilmark{2}}
\affil{Department of Astronomy/Steward Observatory, The University of Arizona, 933 N. Cherry Avenue, Tucson, AZ 85721, USA}
\affil{Lunar and Planetary Laboratory, The University of Arizona, 1629 E. University Boulevard, Tucson, AZ 85721, USA}

\author{Mercedes L\'opez-Morales\altaffilmark{2}}
\affil{Harvard-Smithsonian Center for Astrophysics, 60 Garden Street, Cambridge, MA 01238, USA}

\author{Andr\'es Jord\'an\altaffilmark{2}}
\affil{Instituto de Astrof\'isica, Facultad de F\'isica, Pontificia Universidad Cat\'olica de Chile, Av.\ Vicu\~na Mackenna 4860, 7820436 Macul, Santiago, Chile}
\affil{Millennium Institute of Astrophysics, Chile}

\author{David J. Osip}
\affil{Las Campanas Observatory, Carnegie Institution of Washington, Colina el Pino, Casilla 601 La Serena, Chile}

\author{Nikole K. Lewis}
\affil{Space Telescope Science Institute, 3700 San Martin Drive, Baltimore, MD 21218 USA}

\author{Florian Rodler}
\affil{European Southern Observatory, Alonso de Cordova 3107, Vitacura, Santiago de Chile}
\affil{SETI Institute, 189 Bernardo Ave, Suite 200, Mountain View, CA 94043, USA}

\author{Jonathan D. Fraine}
\affil{Department of Astronomy/Steward Observatory, The University of Arizona, 933 N. Cherry Avenue, Tucson, AZ 85721, USA}

\author{Caroline V. Morley}
\affil{Department of Astronomy and Astrophysics, University of California, Santa Cruz, CA 95064, USA}

\and

\author{Jonathan J. Fortney}
\affil{Department of Astronomy \& Astrophysics, University of California, Santa Cruz, CA 95064, USA}

\altaffiltext{1}{National Science Foundation Graduate Research Fellow.}
\altaffiltext{2}{Earths in Other Solar Systems Team, NASA Nexus for Exoplanet System Science.}


\begin{abstract}

GJ~1214b is the most studied sub-Neptune exoplanet to date. Recent measurements have shown its near-infrared transmission spectrum to be flat, pointing to a high-altitude opacity source in the exoplanet's atmosphere, either equilibrium condensate clouds or photochemical hazes. Many photometric observations have been reported in the optical by different groups, though {simultaneous measurements spanning the entire optical regime are lacking.} We present an optical transmission spectrum (4,500--9,260~\AA) of GJ~1214b in 14 bins measured with Magellan/IMACS repeatedly over three transits. We measure a mean planet-to-star radius ratio of ${R_{p}/R_{s} = 0.1146\pm{2\times10^{-4}}}$ and mean uncertainty of $\sigma(R_{p}/R_{s})=8.7\times10^{-4}$ {in the spectral bins}. The optical transit depths are shallower on average than {observed} in the near-infrared. We present a model for jointly incorporating the effects of a composite photosphere and atmospheric transmission (CPAT) through the exoplanet's limb, and use it to examine the cases of absorber and temperature heterogeneities in the stellar photosphere. We find the optical and near-infrared measurements are best explained by the combination of {(1) photochemical haze in the exoplanetary atmosphere with a mode particle size $r=0.1~\micron$ and haze-forming efficiency $f_{haze}=10 \%$ and (2) faculae in the unocculted stellar disk with a temperature contrast $\Delta T=354^{+46}_{-46}$~K, assuming 3.2\% surface coverage.} The CPAT model can be used to assess potential contributions of heterogeneous stellar photospheres to observations {of exoplanet transmission spectra}, which will be important for searches for spectral features in the optical.

\end{abstract}



\section{INTRODUCTION}

Transmission spectroscopy, in which we study transiting planets at multiple wavelengths, provides \textcolor{black}{a powerful tool for placing} constraints on the nature of close-in exoplanets. The apparent radius of a transiting exoplanet at a given wavelength $R_{p}(\lambda)$ is a function of its atmospheric mean molecular cross section $\sigma(\lambda)$ and scale height $H = \frac{k_{B} T} {\mu g}$, where $k_{B}$ is Boltzmann's constant, $T$ is the temperature, $\mu$ is the atmospheric mean molecular mass, and $g$ is the local gravitational acceleration. Therefore, by examining how an exoplanet blocks the light from its host star at multiple wavelengths, we directly probe both the chemical composition and physical structure of its atmosphere.

\textcolor{black}{In the optical wavelength regime ($\sim0.3$--$1.0~\micron$), this technique provides access to strong atomic lines and molecular bands as well as cloud and haze processes, revealing a diversity of exoplanet atmospheres \citep{Sing:2016aa}. Detections have been reported of Na \RNum{1} \citep{char2002, redf2008, sing2008, sing2012, Sing:2016aa, jens2011, huit2012, zhou2012, niko2014}, K \RNum{1} \citep{sing2011, sing2015a}, and H$_{2}$O \citep{stev2015} in the atmospheres of hot Jupiters. \citet{Evans:2016aa} have presented evidence for TiO/VO in WASP-121b, while non-detections of TiO/VO in other hot, giant exoplanets \citep{huit2013, sing2013} could point to breakdown by stellar activity \citep{knut2010} or the presence of a high-altitude opacity source, either due to lofted cloud decks or photochemical hazes \citep{seag2000, fort2005, howe2012, morl2013, morl2015}. In clear atmospheres, measurements at shorter optical wavelengths directly probe the physics of scattering processes \citep{seag2000, hubb2001, leca2008}, thereby allowing measurements of the atmospheric mean molecular mass \citep{benn2012}. For low-mass transiting exoplanets, this information can provide the key for distinguishing between rocky and gaseous bulk compositions \citep{benn2013}.}

GJ~1214b \citep{char2009} \textcolor{black}{provides an excellent opportunity to study a low-mass planet with transmission spectroscopy}. Orbiting an M4.5 dwarf only \textcolor{black}{14.55 parsecs away \citep{angl2013},} the large transit depth of the planet and apparent brightness of its host star make it very suitable for in-depth study through transmission spectroscopy. Given its relatively small mass (6.3~M$_{\oplus}$) and large radius \cite[2.8~R$_{\oplus}$;][]{angl2013}, the low bulk density of GJ~1214b requires that it contain a substantial gas component, though different admixtures of rock, ice, and volatiles---owing to different formation histories---can explain its bulk density equally well \citep{roge2010}. If GJ~1214b possessed a clear, hydrogen-dominated atmosphere with a $\sim$150--200~km scale height, whether obtained through direct accretion from the protoplanetary nebula or secondary outgassing, absorption features in its transmission spectrum could vary as a function of wavelength by as much as 0.3\% of the host star's flux \citep{mill2010}. Assuming a H-dominated and clear atmosphere, such a planet should in principle produce absorption features detectable by current ground-based and space-based instrumentation.

Attempts to constrain the transmission spectrum of GJ~1214b, however, have revealed a remarkably flat spectrum. Starting with the first results provided by \citet{bean2010}, observations of GJ~1214b in the optical and near-infrared have found a featureless spectrum \citep{bean2011, cros2011, dese2011, bert2012, murg2012, colo2013, demo2013, frai2013, tesk2013, cace2014, gill2014, wils2014, nasc2015}. Reported spectral features in the near-infrared \citep{crol2011, demo2012} have not been reproduced by follow-up measurements in the same bandpasses \citep{nari2013}. Recently, \citet{krei2014} reported the most precise measurements to date obtained during 12 transits with HST/WFC3, which demonstrate a lack of observable features from 1.1--1.7~$\micron$ that rules out cloud-free scenarios for both hydrogen-dominated and high mean molecular mass atmospheres. While exquisite precisions exist for GJ~1214b's transmission spectrum in the near-infrared where its red host star is very bright (H=9.1), its transmission spectrum remains poorly constrained in the blue optical, where the host star is exceedingly faint (B=16.4). The existing optical measurements rely on wide-band photometry and have been compiled from a variety of sources, complicating the detection of spectral features and making the measurements more prone to systematics in the measurement of the transit depth.

Modeling efforts have found that a high-altitude, optically thick layer, whether composed of photochemically produced hydrocarbon hazes \citep{mill2012, howe2012, morl2013, morl2015} or equilibrium condensate clouds \citep{morl2013, morl2015}, can account for the flat near-infrared transmission spectrum by obscuring spectral features that originate lower in the atmosphere. Such a layer could exist for both hydrogen-dominated and high mean molecular mass atmospheres \citep{morl2013}, so the presence of a high-altitude opacity source does not by itself tell us about atmospheric composition. \textcolor{black}{Interestingly, however, the only $V$-band (4,730--6,860~\AA; centered at 5,500~\AA) measurement to date \citep{tesk2013} points to a relatively shallow transit depth, which should be precluded by an high-altitude, optically thick layer.}

Here, we present an optical transmission spectrum \textcolor{black}{(4,500--9,260~\AA)} of GJ~1214b \textcolor{black}{measured} with Magellan/IMACS over three transits, which represents the first transmission spectrum of \textcolor{black}{this sub-Neptune measured simultaneously across the optical wavelength range. We find the transit depths across this range are generally shallower than comparable values in the near-infrared. As no physical model of the exoplanet's atmosphere can reproduce these measurements and the flat near-infrared spectrum simultaneously, we investigate the contribution of a heterogenous stellar photosphere, including faculae and starspots, to the observed transmission spectrum. We find unocculted faculae in the stellar photosphere to be most consistent with the shallower transit depths we observe in the optical.} In Section~\ref{sec:datacollection}, we describe the data collection. We detail the data reduction and our detrending procedure in Sections~\ref{sec:datareduction} and \ref{sec:datadetrending}, respectively, and present our results in Section~\ref{sec:results}.  We discuss the physical interpretation of the spectrum in Section~\ref{sec:discussion}, \textcolor{black}{presenting a model for incorporating the effects of a composite photosphere and atmospheric transmission through the exoplanet's limb and applying it to the cases of absorber and temperature heterogeneities in the stellar photosphere. We summarize our findings and their implications in Section~\ref{sec:summary}.}

\section{DATA COLLECTION \label{sec:datacollection}}

\textcolor{black}{The observations of GJ~1214b were collected as part of {\em ACCESS}, the Arizona-CfA-C\'atolica Exoplanet Spectroscopy Survey. In this section we first give a brief summary of {\em ACCESS} before describing the specific observations of GJ~1214b in detail.}

\subsection{\textcolor{black}{\em ACCESS}}

{\em ACCESS} \textcolor{black}{is a collaborative project between the University of Arizona, the Harvard-Smithsonian Center for Astrophysics, the Pontificia Universidad Cat\'olica de Chile, and the Carnegie Institution for Science with the aim of measuring optical transmission spectra from a representative sample of transiting exoplanets}. Our targets \textcolor{black}{include 30 planets with masses and radii between 6--450~M$_{\oplus}$ and 2.5--23~R$_{\oplus}$, and effective temperatures ($T_{eff}$) between 600--2,800~K}. 
\textcolor{black}{\textit{ACCESS} utilizes} ground-based, multi-object spectrographs \textcolor{black}{(MOS)} to simultaneously collect spectra from the exoplanet host star and many comparison stars in the same field of view, \textcolor{black}{enabling corrections for} systematic noise sources arising from the instrument or variable weather conditions. Our survey design emphasizes repeated observations to ensure the reliability of our findings. \textcolor{black}{We demonstrated the feasibility of this technique in a pilot study on WASP-6b \citep{jord2013}, which precisely measured the optical transmission spectrum of that transiting \textcolor{black}{hot Jupiter} in a single transit with Magellan/IMACS despite strongly variable transparency during part of the transit. The resulting spectrum for WASP-6b was most consistent with scattering, a result confirmed later by HST \citep{niko2015, Sing:2016aa}} 

\subsection{\textcolor{black}{\em Observational Design}}

We observed three transits of GJ~1214b with \textcolor{black}{the Inamori-Magellan Areal Camera \& Spectrograph (IMACS), a versatile wide-field imager and spectrograph permanently mounted on the Magellan Baade Telescope. It includes two cameras, one at each of the f/2 and f/4 foci, which can each be used in} \textcolor{black}{MOS} \textcolor{black}{mode with custom-designed slit masks. The f/2 camera covers a $27'$ diameter circular field and provides spectra with resolving powers up to $R\sim1,200$. The f/4 provides higher resolution spectra, up to $R\sim5,000$, over a smaller, $15' $ square field. The detector of each camera is comprised of eight $2K\times4K$~CCDs, forming an $8K\times8K$ mosaic \citep{Dressler:2006aa}.}

\textcolor{black}{For these observations we used in turn each of the IMACS cameras in} \textcolor{black}{MOS} \textcolor{black}{mode. We observed two transits} with the f/4 camera and one with the f/2 camera; therefore we designed two masks, one for each of the cameras. In the mask design we considered three criteria: 1) to include the widest spectral range for the target and comparison stars; 2) to include as many comparison stars as possible; and 3) to eliminate slit losses using extra-wide slits ($5''$ for the f/4 mask and $10''$ for the f/2 mask). We also used larger lengths for the slits ($12''$ for the f/4 mask and $22''$ for the f/2 mask) in order to adequately sample the sky background.

We selected comparison stars using multiple criteria. Starting from an initial list of all stars in the UCAC4 catalog \citep{zach2013} in the same field of view as the target, we eliminated all known binary stars.  We then made magnitude cuts, including only stars less than 0.5 magnitudes brighter and 1 magnitude fainter than the target in V band.  Finally, we prioritized the remaining candidates by their distance ($D$) from the target star in B$-$V / J$-$K 
-color space,
\begin{equation}
D = \sqrt{ {[(B-V)_{c} - (B-V)_{t} ]}^{2} + {[ (J-K)_{c} - (J-K)_{t} ]}^{2} },
\end{equation}
in which the subscripts $c$ and $t$ refer to the comparison and target stars, respectively, with the closest stars in this parameter space receiving the highest priorities. \textcolor{black}{This ranking is especially important in the case of GJ~1214b because the host star is much redder than other stars in its field.} Using these rankings, we adjusted the pointing offset and rotation of the IMACS mask to maximize the wavelength coverage for the target and a maximum number of comparison stars. Our final configurations allowed us to cut slits for the target and 14 comparison stars on the f/4 mask and 24 comparison stars on the wider-field f/2 mask. Finally, we included 25 $5\times5''$ square slits in each mask matching the position of relatively faint stars in the field. Those boxes are used to align the masks.

\subsection{\em Observations}

Details for the three transit observations are provided in Table~\ref{table:log}. We carried out all observations in spectroscopic mode with no filter. \textcolor{black}{We utilized $2\times2$ binning to reduce read-out times, and chose i}ntegration times to provide a maximum of roughly 25,000--30,000 counts \textcolor{black}{in analog-to-digital units (ADU; gain~$=0.56~e^{-1}/$ADU for f/4 setup, $1.0~e^{-1}/$ADU for f/2 setup)} \textcolor{black}{per resolution element} on the target spectrum, which was the brightest of the spectra. A sample spectrum for the target is shown in Figure~\ref{fig:data_example}. For Transits 1 and 2, we used the 150 line/mm grating, which provided usable coverage of 4,500--9,260~\AA\ on the target spectrum with a chip gap from 7,054--7,254~\AA~(Figure~\ref{fig:spectra}). \textcolor{black}{Given the wide slits used, the spectral resolution for each exposure was set by the seeing (see Table~\ref{table:log}), with the average being $R\sim240$}. For Transit~3, we utilized the f/2 camera for two reasons: 1) combining results from two different cameras provides an additional level of cross-validation for the resulting transmission spectrum, and 2) the larger field of view for the f/2 camera allowed us to obtain more comparison spectra from potentially better comparison stars. We used the 300 line/mm grism (blazed at 17.5 degrees), which gave spectral coverage of 4,500--9,260~\AA~on the target spectrum with a chip gap from 6,506--6,596~\AA~(Figure~\ref{fig:spectra}). With $2\times2$ binning and variable seeing ($0.5-1.0''$), the \textcolor{black}{spectral} resolution for each exposure varied between \textcolor{black}{$R\sim300$--700, with an average of $R\sim480$}.

\textcolor{black}{In each night} we collected bias frames, dark frames, and quartz lamp flat frames with the science mask and the same setup used for the science observations. The bias frames demonstrated that the bias levels are essentially constant across the detector, so we adopted constant bias levels using the median of the overscan region on each science frame. The dark frames showed the dark count to be negligible for the exposure times used, so no dark subtraction was applied. 
We reduced the data with and without flat field corrections, and found the detrended light curves of GJ~1214 with the f/4 setup (Transits~1 and 2) displayed more correlated noise when flat-fielded. Flat-fielding the f/2 data (Transit~3), however, reduced the correlated noise contribution to the final light curve. More precisely, applying a flat field correction increased the variance of the flicker noise models (see Section~\ref{sec:datadetrending}) for Transits~1 and 2 by 4\% and 22\%, respectively, and decreased it for Transit~3 by 21\%. For the rest of this analysis, we used the non-flat-fielded spectra for Transits~1 and 2 (f/4 camera) and the flat-fielded spectra for Transit~3 (f/2 camera). To obtain a wavelength solution, we took exposures of He, Ne, and Ar lamps before and after the science observations using calibration masks identical to the science masks but with $0.5''$ slits.

\section{DATA REDUCTION \label{sec:datareduction}}

We reduced each dataset with a custom, Python-based pipeline following the procedure employed in an earlier analysis of WASP-6b \citep{jord2013}.  Below, we summarize the key steps in the reduction procedure. 

\subsection{\em Spectrum tracing}

Bias levels were estimated and removed for each integration using the median of the overscan region of each chip.  For each frame, the position of each spectrum was then traced by identifying the centroid of the spectrum's spatial profile for each resolution element ($2\times2$ binned pixel) in the dispersion or spectral direction, and robustly fitting a ${2^{nd}}$ order polynomial to the identified centroids (Figure~\ref{fig:data_example}), taking into account chip gaps when necessary. The left and right slit borders were identified using \textcolor{black}{the SciPy\footnote{\url{http://www.scipy.org/}} implementation of} a Prewitt filter, which approximates the spatial gradient of images and produces maximal values at edges. In this case, the Prewitt-filtered images display strong positive values at slit borders. Traces of the slit borders were obtained by robustly fitting polynomials to the positions of the maxima of the Prewitt-filtered images flanking the spectral trace.

\subsection{\em \textcolor{black}{S}ky Subtraction}

The sky spectrum was identified for each resolution element in the spectral direction using the median of {all} resolution elements within the slit borders, excluding {an} aperture {centered on the spectral trace containing} the stellar spectrum. The {central} aperture sizes were selected to minimize the correlated noise contribution to the final light curve. {Central a}pertures of 20 ($4.4''$), 24 ($5.3''$), and 24 ($9.6''$) resolution elements were used for the Transits~1, 2, and 3 datasets, respectively.  The sky spectrum \textcolor{black}{was} subtracted from the signal within the central aperture, leaving only the profile of the stellar spectrum. The final extracted spectrum was then obtained by summing the spectrum profile within the central aperture in the spatial direction. Optimal extraction \citep{mars1989} did not give noticeable gains over the simple extraction for our high signal-to-noise spectra.

\subsection{\em Wavelength Calibration}

Given the wide slits used, skylines recorded during the observations had too low of a spectral resolution to be useful for wavelength calibration.  Therefore, the arc lamps taken before and after the science observations with the narrow-slit ($0.5''$) calibration mask were used to calibrate the extracted spectra. Lorentzian profiles were fitted to each spectral line to determine their centroids in the dispersion direction. Using these pixel positions and the known vacuum wavelengths, the wavelength solution for each spectrum was found by an iterative process in which a $6^{th}$ order polynomial was fitted to the wavelengths as a function of pixel position, the data point with the greatest deviation from the fit was removed, and the process was repeated until the root mean square error value of the fit was less than 2000~m/s (${\sim}0.05~${\AA}).  Depending on the wavelength coverage of the particular spectrum, between 50--60 spectral lines were assigned a pixel position, and roughly 35--45 were utilized in the final fit.

The wavelength solution found with the arc lamps was used for the first science spectrum of the night, and the remaining science spectra were cross-correlated with the first to determine their respective wavelength shifts.  \textcolor{black}{As a function of time, the positions of the spectra drifted slowly in the dispersion direction}, so a $3^{rd}$ order polynomial was fitted to the shifts identified via the cross-correlations, and the ``smoothed'' wavelength shifts provided by this fit were used to interpolate all spectra into a common wavelength grid using b-splines.  This step removed wavelength shifts of roughly 10~{\AA} between spectra over the course of the night. To identify any residual wavelengths shift on the order of 1~{\AA}, the Fourier transform of each spectrum was \textcolor{black}{multiplied by} the Fourier transform of the median spectrum, and the peak of the convolution function was used to identify the remaining wavelength shift required.

The above steps ensured that all spectra from a single star were calibrated to the same reference frame.  As a final step, the spectra from all stars were calibrated to the same physical reference frame by identifying shifts between the H-alpha absorption line minimum of the median spectra and the vacuum wavelength of H-alpha, and then interpolating the spectra onto a common wavelength grid using b-splines.  For GJ~1214, in which the H-alpha line was not evident, the Na 8,200~\AA\ doublet minimum was used for this process instead.

\section{DATA DETRENDING \label{sec:datadetrending}}

\textcolor{black}{After extracting and wavelength-calibrating the spectra, we generated sets of light curves for GJ~1214 and the comparison stars to identify each transit and related systematics. We first generated white light curves, integrating all the light between 4,500--9,260~\AA~in each spectrum. We also generated spectroscopically resolved light curves using 14 bins with varying widths that allowed for similar precisions on the light curve parameters between bins (see~Section~\ref{subsec:spectroscopic}). All these light curves were systematics-dominated. We developed the formalisms described below to model out those systematics.}

\subsection{\em Signal Modeling Frameworks}

We employed two modeling frameworks to understand and remove systematic trends from the light curve of GJ~1214\textcolor{black}{, which we label} the ``PCA-based'' and the ``polynomial-based''  frameworks. 

\subsubsection{\em PCA-based modeling framework}

Following the framework detailed by \citet{jord2013}, we modeled the observed target light curve $l(t)$ as 
\begin{equation}
l(t) = F T(t, \theta) S(t)  \epsilon(t),
\end{equation}
in which \textcolor{black}{$t$ is time,} $F$ is the underlying flux from the target star, $T(t,\theta)$ is the transit signal defined by the vector of transit parameters $\theta$, $S(t)$ is the perturbation signal due to the combination of all systematic variations, and $\epsilon(t)$ is the stochastic noise component. We account for both uncorrelated variations (``white noise'') and correlated variations (``red noise'') in the light curve with the $\epsilon(t)$ term and modeled them following the wavelet-based method of \citet{cart2009}\footnote{A Python implementation of this procedure can be found at \url{http://www.github.com/nespinoza/flicker-noise}.} (see Section~\ref{subsubsec:MCMC}). We assumed the underlying flux from the target star to be constant, so any actual variations in the star's flux were therefore encapsulated by the noise component. We assumed the perturbation signal to be a linear combination of signals owing to different instrumental and atmospheric effects, which can be represented mathematically as 
\begin{equation}
S(t) = \prod\limits_{i=0}^{n} \alpha_{i}s_{i}(t),
\end{equation}
in which $s_{i}(t)$ represents the different signals and $\alpha_{i}$ their respective scaling coefficients. Expressed in logarithmic space, each of these multiplicative signals become additive, so defining $L \equiv \log{l(t)}$, the base-10 logarithm of the observed target light curve can be written as
\begin{equation}
L = \log{F} + \log{T(t, \theta)} + \sum\limits_{i=0}^{n} \alpha_{i}s_{i}(t) + \log{\epsilon(t)}.
\label{eq:L}
\end{equation}
The systematic variations represented by the perturbation signal include common variations experienced by the target and comparison spectra due to instrumental and atmospheric effects. Therefore, within this framework the observed signal from each $k$ comparison star can be modeled as
\begin{equation}
L_{k} = \log{F_{k}} + \sum\limits_{i=0}^{n} \alpha_{i,k}s_{i}(t) + \log{\epsilon_{k}(t)},
\label{eq:L_k}
\end{equation}
in which $F_{k}$ is the baseline flux from the comparison star, $\alpha_{k}$ is the set of unique scaling coefficients for the systematic variations as they apply to this comparison star, and $\epsilon_{k}(t)$ is the unique noise signal associated with this light curve. Thus, the mean-subtracted light curve from each comparison star ($L_{k} - \log{F_{k}}$) can be used to estimate the perturbation signal $S(t)$, which can then be subtracted from $L$ in Eq.~\ref{eq:L}.

We performed a principal component analysis (PCA) of the mean-subtracted comparison light curves to estimate the independent signals $s_{i}(t)$ comprising the perturbation signal $S(t)$. With $N$ comparison stars, we could estimate at most $N$ principal components via PCA. As detailed in Section~\ref{subsubsec:MCMC}, we let the scaling coefficients $\alpha_{i}$ and the baseline flux from the host star $F$ float as free parameters in the Markov chain Monte Carlo procedure, so we enforced the maximum number of principal components for use in the model to be $M \le N -1 $. However, we determined the optimal number of components to be the minimum number that could achieve the same predictive power as the best set of $M$ available components.

We estimated the predictive power of each set of components using a $k$-fold cross-validation procedure. First, we split the out-of-transit target light curve into 20 segments (or ``folds''). For each number of available principal components (from 1 to $M$), we fit the $\alpha$ scaling coefficients for the available components to the target light curve using 19 ``training'' folds, and recorded the error (in units of normalized flux) between the target light curve and fit for the single ``validation'' fold. Repeating this process using each of the 20 folds in turn as a validation fold gave us an estimate and confidence interval of the prediction error for each number of principal components. We identified the number of principal components that produced the smallest median prediction error (``the best set''), and then used the lowest number of principal components with a median prediction error indistinguishable at the $1\sigma$ level from that of the best set.

The final consideration in the PCA-based procedure was to select the comparison stars to ultimately use in generating the principal components. Each comparison light curve provides a noisy estimate of the perturbation signal, so it follows that the combination of comparison stars can be optimized to provide the best estimate with the least noise. Through successive iterations, we found that for each night the brightest comparison stars produced principal components that were most capable of accurately predicting perturbations in the target light curve. Ultimately, we used the brightest 4, 5, and 4 comparison stars to estimate the perturbation signals for the Transits 1, 2, and 3, respectively. We found these comparison stars to also provide the best results when utilizing the polynomial-based modeling framework we explored.

\subsubsection{\em Polynomial-based modeling framework}

We also modeled the observed target light curve following the procedure employed by \citet{bean2010} in their successful VLT/FORS observations of GJ~1214b. This framework is empirically motivated, as the data show that simply dividing the target light curve by the sum of the comparison light curves removes most of the variations in the out-of-transit flux from the target, leaving only a smoothly varying long-term trend. Like \citet{bean2010}, we find that this trend can be modeled well as a 2$^{nd}$ order polynomial function of time, and attribute it to the color difference between this very red target star and the available comparison stars \textcolor{black}{in the field}. In this framework, the target light curve can be expressed as 
\begin{equation}
l(t) = F T(t, \theta) S(t) \sum\limits_{i=0}^{2} \alpha_{i}t^{i} \epsilon(t),
\end{equation}
and the comparison light curves as
\begin{equation}
l_{k}(t) = F_{k} S(t) \epsilon_{k}(t),
\end{equation}
in which $\alpha_{i}$ now describes the polynomial coefficients and the other terms have the same meaning as above. Dividing by the sum of comparison light curves, the detrended target light curve can be expressed as
\begin{equation}
l_{det}(t) = T(t, \theta) \sum\limits_{i=0}^{2} \alpha_{i}t^{i} \epsilon_c(t),
\end{equation}
in which the noise term $\epsilon_c(t)$ now represents the combined noise from the target and comparison light curves, and the constant $F$ and $F_{k}$ terms have been subsumed into the $\alpha_{0}$ coefficient.
{We note that this formalism is an approximation, since $S(t)$ will be different for the target and reference stars in real data (as expressed by the $\alpha_{i,k}$ coefficients in Eq.~\ref{eq:L_k}) and will not divide out exactly.}

Both modeling frameworks are tested in the following.  We {compare their effectiveness in Section~\ref{subsec:comparing_methods} and} report the results of the polynominal-based detrending procedure in this paper.

\subsection{\em Markov Chain Monte Carlo Procedure \label{subsec:MCMC}}
\subsubsection{General procedure \label{subsubsec:MCMC}}

For both modeling frameworks, the transit parameters were estimated for each night using a Markov chain Monte Carlo (MCMC) optimization procedure.  The transit signal was modeled following the formalism of Mandel {\&} Agol (2002), which accounts for the effect of limb darkening via a quadratic law of the form
\begin{equation}
I(\mu) = I(1)[1 - u_{1}(1- \mu) - u_{2}(1 - \mu)^{2}],
\end{equation}
in which $\mu$ is the cosine of the angle between the stellar surface normal and the line of sight to the observer and $u_{1}$ and $u_{2}$ are the limb darkening coefficients. \textcolor{black}{Fixing the limb darkening coefficients has been shown to bias measurements of the transit depth \citep{espi2015}. Yet, when left as free parameters, these coefficients are strongly correlated in MCMC retrievals \citep{pal2008, kipp2013}. However, following} a rotation onto new principal axes:
\begin{mathletters}
\begin{eqnarray}
\omega_{1} = u_{1}\cos{\phi} - u_{2}\sin{\phi}, \\
\omega_{2} = u_{2}\cos{\phi} + u_{1}\sin{\phi}
\end{eqnarray}
\label{eq:rotation}
\end{mathletters}
with $\phi=35.8\degree$ \citep{kipp2013}, $\omega_{1}$ and $\omega_{2}$ are essentially uncorrelated, and the first parameter can account for variations induced by the transit geometry while the second remains constant \citep{howa2011}. Therefore, we used these rotated coefficients to describe the effect of limb darkening, leaving $\omega_{1}$ as a free parameter and fixing $\omega_{2}$ to values obtained from a PHOENIX atmospheric model \citep{huss2013} for $\mu\ge0.1$ with \textcolor{black}{stellar parameters closest to those identified by \citet{rojas2012}: $T_{eff}=3,300~$K, $\log{g}=5.0$, and $[M/H]=0.0$}. We used a uniform prior on $\omega_{1}$ with boundary values set by the triangular sampling method of \citet{kipp2013}.

In each MCMC procedure, we adopted the fixed parameters on the transit model, which included system scale ($a/R_{s}$), inclination ($i$), orbital period ($P$), eccentricity ($e$), and argument of periastron ($\omega$), from \textcolor{black}{\citet{krei2014} to allow for direct comparisons between results}. The planet-to-star radius ratio ($R_{p}/R_{s}$), rotated limb darkening coefficient ($\omega_{1}$), scaling coefficients for principal components or polynomial terms ($\alpha$), and photometric uncertainty ($\epsilon$) were left as free parameters. The baseline flux ($F$) was also left as a free parameter in the PCA-based procedure. We placed a Gaussian prior on $R_{p}/R_{s}$, centered on the \textcolor{black}{median transit depth reported by \citet{krei2014}, with an uncertainty of $\sigma_{R_{p}/R_{s}}=0.01$} to allow the algorithm to thoroughly explore the parameter space, and truncated by the range [0,1]. We also placed $5\sigma$ Gaussian priors on each $\alpha$ coefficient. In the PCA-based procedure, we determined the mean and standard deviation for the priors using the distributions of the $\alpha$ coefficients obtained for the comparison stars. In the polynomial-based procedure, we performed a bootstrap analysis on the out-of-transit data to determine the mean and standard deviation for the priors. We resampled the out-of-transit data with replacement 1,000 times and fit a $2^{nd}$ order polynomial function of time to each sample via least-squares, recording the fit coefficients. We utilized the mean and standard deviation of the coefficient distributions to determine the $5\sigma$ Gaussian priors for the MCMC. Finally, we placed a $5\sigma$ Gaussian prior on $F$ in the PCA-based procedure using the median and standard error of the out-of-transit flux.  

We ran five chains of 130,000 steps and discarded the first 30,000 steps as the burn-in. For each step in the chain, the likelihood of the residuals given the model was calculated via the same likelihood function given by Equation 41 of \citet{cart2009}, which parameterizes the contributions of uncorrelated and time-correlated noise sources to the observed light curve via the parameters $\sigma_{w}$ and $\sigma_{r}$, respectively. We placed uniform priors bounded by the interval [0,1] on both of these parameters. 

We combined the results from all chains to determine the posterior values. We evaluated convergence between chains using the Gelman-Rubin statistic $\hat{R}$ \citep{gelm1992}, and considered the chains to be well-mixed if $\hat{R} \leq 1.03$ for all parameters. We constructed the posterior distribution by sampling each chain at intervals spaced by $10\times$~the half-life of the autocorrelation in the chain. This wide spacing ensured independent sampling. We adopted the median and \textcolor{black}{68.27\% confidence interval defined by the 15.87$^{th}$ and 84.14$^{th}$} percentiles of the posterior distribution as the final posterior value and uncertainty for each parameter.

\subsubsection{White-light light curve analysis}

We first analyzed the white light curve, using flux from the complete spectra (4,500--9,260~\AA) of GJ~1214 and the comparison stars. We performed the procedure as outlined above, additionally including the time of mid-transit ($t_{0}$) as a free parameter. We placed a uniform prior on $t_{0}$ spanning the complete observation.

\subsubsection{Spectroscopic light curve analysis \label{subsec:spectroscopic}}

Following the white light analysis, we repeated the MCMC procedure for each wavelength bin.  We allowed the same parameters to float, except $t_{0}$, which we fixed to the value obtained from the white light analysis.

\textcolor{black}{In determining the width of wavelength bins for the spectroscopic analysis, we aimed to 1) maximize the number of bins, while 2) collecting enough signal in the bins to discriminate between model transmission spectra with signals on the order of $\Delta (R_{p}/R_{s}) \sim 0.005$. Taking into account our usable wavelength coverage of 4,500--9,260~\AA\ and the chip gaps for the f/4 and f/2 setups, we determined that dividing the spectra into 14 bins with roughly even signal to be optimal: we use 10 bins with a mean width of $~200$~\AA~(174--243~\AA) at wavelengths greater than $7,250$~\AA, where the spectra are brightest, and four bins with a mean width of $616$~\AA~(310--1,157~\AA) at shorter wavelengths, where the spectra are considerably fainter (Table~\ref{table:rprs}).}

\subsubsection{Other limb darkening prescriptions}

Along with the limb darkening prescription described above we investigated two additional methods for handling the effect of stellar limb darkening on the transit light curve. In the first approach we fixed the limb darkening coefficients $\omega_{1}$ and $\omega_{2}$ during the MCMC analysis to the values from the PHOENIX atmospheric model \citep{huss2013}. In the second we utilized the triangular sampling method proposed by \citet{kipp2013}, which involves fitting for new limb darkening coefficients $q_{1} = \left(u_{1} + u_{2} \right)^{2}$ and $q_{2} = 0.5u_{1}\left(u_{1} + u_{2} \right)^{-1}$ with uniform priors on the interval [0,1] in order to efficiently sample the physically bounded parameter space. The $R_{p}/R_{s}$ values for the final transmission spectrum did not differ significantly between the three limb darkening prescriptions. With respect to the nominal results, adopting fixed limb darkening coefficients caused the $R_{p}/R_{s}$ values on average to differ by $5\times10^{-4}$ ($0.6\sigma$). Similarly, the triangular sampling method led to $R_{p}/R_{s}$ values that differed by $3\times10^{-4}$ ($0.3\sigma$). Both methods enlarged the $1\sigma$ uncertainties on the transmission spectrum by $0.2\%$. In this paper we report the $R_{p}/R_{s}$ values determined via our nominal limb darkening prescription described in Section \ref{subsubsec:MCMC}.

\section{RESULTS \label{sec:results}}

\subsection{\em Comparison of the Detrending Methods \label{subsec:comparing_methods}}

\textcolor{black}{We find the polynomial-based detrending procedure to be the most effective at removing long-term trends in the light curves.  By contrast, strong long-term trends in the light curves of GJ~1214 remain after applying the PCA-based procedure. We attribute these to effects introduced by the color difference between the target and comparison stars, which are not encapsulated in the principal components of the comparison light curves. As the wavelength-based likelihood function we employ \citep{cart2009} is particularly sensitive to time-correlated systematics, this results in inflated parameter uncertainties. Accordingly, the uncertainties on $R_{p}/R_{s}$ we find with the PCA-based procedure are on average $1.4\%$ larger than those from the polynomial-based procedure. The final $R_{p}/R_{s}$ values from the PCA-based procedure differ on average from those of the polynomial-based procedure by $1.6\times10^{-3}$ ($1.3\sigma$). We conducted the analysis described in Section~\ref{sec:discussion} on the white-light and spectroscopic light curves extracted from both detrending procedures. The choice of detrending procedure does not affect the interpretation of the spectrum or the conclusions of this paper, though the inflated uncertainties from the PCA-based procedure lead to looser constraints.  Given 1) the consistency between detrending methods, 2) the potential for time-correlated systematics to bias the measured transit depths, and 3) the tighter constraints provided by the polynomial-based detrending procedure, we report the results from the polynomial-based procedure and adopt them for the white-light and spectroscopic analyses.}

\subsection{\em White-light Light Curve Fitting \label{subsec:whiteLC}}

The $t_{0}$ values determined via the polynomial-based and PCA-based detrending procedures {agree} within $0.5\sigma$. We report the mid-transit times from the polynomial-based method in Table~\ref{table:t0} and use them for the remainder of this work. Figure \ref{fig:wl} displays the white light data for each night and their best-fitting transit models.

\subsection{\em Spectroscopic Light Curve Fitting \label{subsec:specLC}}

We show the spectroscopic light curves obtained via the polynomial-based procedure for Transits 1, 2, and 3, respectively, with their best-fitting transit models and residuals in Figures \ref{fig:t1poly}--\ref{fig:t3poly}. The median residual RMS values for the spectroscopic light curves are $1.5\times$, $2.2\times$, and $3.3\times$~the {photon} noise limit for Transits 1, 2, and 3, respectively. The spectroscopic $R_{p}/R_{s}$ values for each transit are listed in Table~\ref{table:rprs} and shown in Figure~\ref{fig:tspoly}.

\subsection{\em Combining Results \label{subsec:combining_results}}

We {take} the weighted mean of the transmission spectra from the three transits to generate a final transmission spectrum (Table~\ref{table:rprs}). For each wavelength bin, we {average} the $R_{p}/R_{s}$ measurements from each transit {weighted} by their inverse variances {and calculate the uncertainty as} the square {root} of the weighted sample variance. The final combined spectrum is shown in black circles in Figure~\ref{fig:tspoly}.

\section{DISCUSSION \label{sec:discussion}}

\subsection{\em Previous Measurements \label{subsec:previous_measurements}}

Since its discovery, many groups have published measurements of GJ~1214b's transmission spectrum in the optical and near-infrared \citep{bean2010, bean2011, crol2011, cros2011, dese2011, bert2012, demo2012, murg2012, colo2013, demo2013, frai2013, nari2013, tesk2013, cace2014, krei2014, gill2014, wils2014, nasc2015}. 
{A total of 58 measurements have been published for bandpasses with central wavelengths shorter than 9,260~\AA, the longest wavelength in our Magellan/IMACS spectrum. These measurements tend \textbf{to} lie above our data and have a mean of ${R_{p}/R_{s}=0.1170 \pm 2 \times 10^{-4}}$, in which the quoted uncertainty is the standard error of the mean. However, they span a notable range, with a minimum of ${R_{p}/R_{s}=0.1104 \pm 0.0014}$ at 8,550~\AA~\citep{wils2014}, a maximum of ${R_{p}/R_{s}=0.1217 \pm 0.0025}$ at 6,560~\AA~\citep{murg2012}, and a standard deviation of ${1.8 \times 10^{-3}}$.}
The large body of work on GJ~1214b reflects a diverse set of approaches, which can {complicate the combined analysis of data from different sources}. A uniform reanalysis of all existing GJ~1214b observations would be fruitful but is outside the scope of this work.

Of the existing measurements, the 1.1--1.7~$\micron$ spectrum from \citet{krei2014}, obtained during 12 transits with HST/WFC3, provides the most precise measurements and places the tightest constraints on the nature of GJ~1214b's atmosphere.  We adopt the system parameters ($a/R_{s}=15.23$, $i=89.1^{\circ}$, $P=1.58040464894$~days, $e=0$), from \citet{krei2014} in this study to facilitate direct comparisons with those results.

While fitting model transmission spectra, we perform a simultaneous fit of the new 0.45--0.93~$\micron$ Magellan/IMACS spectrum and two highly constraining published datasets: the 1.1--1.7~$\micron$ spectrum from \citet{krei2014} and the Spitzer 3.6 and 4.5~$\micron$ bands from \citet{frai2013}. We adopt the \citet{frai2013} results found using the \citet{bert2012} system parameters, which are the most similar to the \citet{krei2014} parameters.

\subsection{\em Comparison to Model Transmission Spectra \label{subsec:comparison_model_ts}}

We compare the joint dataset to model transmission spectra from \citet{morl2015} that include either KCl and ZnS equilibrium condensate clouds or photochemically produced hydrocarbon hazes. In terms of equilibrium clouds, we consider a grid of 24 models with a range of metallicities (100--1000$\times$~solar) and cloud thicknesses, parameterized by the sedimentation efficiency $f_{sed}$, which is the ratio of the sedimentation velocity to the convective velocity. We consider cloud-free models (no $f_{sed}$ value) and a range of models with thinner ($f_{sed}=1$) to thicker ($f_{sed}=0.01$) cloud layers. With respect to photochemical hazes, we consider a grid of 20 models with vertical eddy diffusion coefficients of $K_{zz}=10^{10}~\textrm{cm}^{2}~\textrm{s}^{-1}$, and a range of mode particle radii $r$ (0.01--1~$\micron$) and haze-forming efficiencies $f_{haze}$ (1--30\%), which represent the mass fraction of precursors that form soots.  All models were calculated at 1$\times$~GJ~1214b's incident stellar flux.

We investigate the goodness-of-fit of the data to the models using the $\chi^{2}$ statistic and represent the results as in \citet{morl2015}. In fitting the models to the data, we allow for a uniform offset in $R_{p}/R_{s}$ to minimize the $\chi^{2}$ value. We assume 37 degrees of freedom ($DOF$; 38 data points $-$ 1 fitted parameter) when calculating the reduced $\chi^{2}$ statistic ($\chi^{2}_{red}$). Tables~\ref{table:X2_simple_cloud} and \ref{table:X2_simple_haze} give the results for the cloud and haze models, respectively. For cloudy atmospheres, models with higher metallicities and thicker clouds ($f_{sed}=\hbox{0.01--0.1}$) tend to provide better fits (Figure~\ref{fig:X2_simple}). The best-fitting cloud model has the highest metallicity and relatively thick clouds ($1000\times$~solar, $f_{sed}=0.1$, $\chi^{2}=81.3$, $\chi^{2}_{red}=2.20$).  
For hazy atmospheres, the models with higher haze-forming efficiencies ($f_{haze}=\hbox{10--30}\%$) and larger mode particle sizes ($r=\hbox{0.3--1}~\micron$) provide the best fits (Figure~\ref{fig:X2_simple}). The best-fitting haze model includes a relatively large mode particle size and haze-forming efficiency ($r=0.3~\micron$, $f_{haze}=10\%$, $\chi^{2}=74.1$, $\chi^{2}_{red}=2.00$).
{A flat line model (1 fitted parameter; $DOF=37$) provides a fit to the data comparable to that of the best-fitting cloud and haze models ($R_{p}/R_{s}=0.11613$, $\chi^{2}=75.2$, $\chi^{2}_{red}=2.03$).}

Given the high precisions of the near-infrared measurements, the model fits are driven by these data. For both the equilibrium cloud and photochemical haze grids, the best-fitting models are those that could most effectively flatten the planet's near-infrared transmission spectrum in agreement with the observations. However, these models are inconsistent with the optical data (Figure \ref{fig:ts_modelfits_simple}). In effect, the large opacity source required to obscure spectral features in the near-infrared predicts an optical spectrum that is either flat and in line with the near-infrared measurements or slightly increasing with shorter wavelengths. Yet the optical data \textcolor{black}{are} in-fact offset below the near-infrared data, with a mean value of $R_{p}/R_{s} = 0.1146\pm{2\times10^{-4}}$ compared to $R_{p}/R_{s} = 0.11615\pm{3\times10^{-5}}$ for the \citet{krei2014} dataset, in which the quoted uncertainty is the standard error of the mean. Thus, the mean of the optical data points is $8\sigma$ lower than that of the near-infrared data. 

To evaluate the significance of the offset between our Magellan/IMACS data and the HST/WFC3 data, we calculate the $\chi^{2}$ fit of our optical data to a flat transmission spectrum given by the mean of the {HST/WFC3 data}. With 14 data points and 1 fitted parameter, we assume 13 $DOF$ when calculating the $\chi^{2}_{red}$ statistic. We find the optical data are inconsistent with the {mean $R_{p}/R_{s}$ of} the HST/WFC3 data at high significance ($\chi^{2}=54.2$, $\chi^{2}_{red}=4.17$, $p<1\times10^{-5}$). Since atmospheric models that are consistent with the flat near-infrared spectrum predict an optical spectrum that is in line with or slightly elevated with respect to the near-infrared, we conclude that, by taking into account transmission through the exoplanet's atmosphere alone, none of the physically plausible models we considered can reproduce both the optical and near-infrared measurements.  With that in mind, we consider in the following sections possible contributions from the star's photosphere to the observed transmission spectrum.

\subsection{\em Effects of a Heterogenous Stellar Photosphere \label{subsec:CPAT}}

\subsubsection{Composite Photosphere and Atmospheric Transmission (CPAT) model}

We investigate how heterogeneities across the star's photosphere could affect the observed transmission spectrum of the exoplanet using a basic model to incorporate the effects of a composite photosphere and atmospheric transmission along the planet's limb (hereafter, the ``CPAT model''). We consider the simplest case in which the emergent spectrum of the star is composed of two distinct components:  the \textcolor{black}{spectrum typical of the occulted transit chord $S_{o}$ (the ``occulted'' spectrum)} and the unocculted spectrum $S_{u}$, which is fully outside of the transit chord (Figure~\ref{fig:schematic}).  It is important to note that the occulted spectrum includes the transit chord but is not limited to it. For example, the planet could transit a region with a spectrum that is typical of $80\%$ of the photosphere, while another $20\%$ of the photosphere produces a distinct spectrum that is not probed by the transit chord.  Additionally, the unocculted region does not need to be continuous in this model. We note that the planet could transit multiple regions with distinct spectral characteristics \textcolor{black}{\citep[as is the case while crossing starspots, e.g.,][]{sanc2011}}, or multiple unocculted regions with distinct spectra could exist, though we show below that this simple model is sufficient to reproduce the observations.

In the case that the stellar photosphere is composed of two distinct components, the observed transmission spectrum given by the CPAT model is
\begin{equation}
\left( \frac{R_{p}}{R_{s}} \right)_{\lambda, obs} = \sqrt{ 1 - \frac{(1-F-D_{\lambda}) S_{o} + F S_{u}}{(1-F) S_{o} + F S_{u}} },
\label{eq:CPAT}
\end{equation}
in which $({R}_{p}/R_{s})_{\lambda, obs}$ is the observed, wavelength-dependent planet-to-star radius ratio, $F$ is the fraction of the stellar disk covered by the unocculted spectrum $S_{u}$, and $D_{\lambda}=({R}_{p}/R_{s})^{2}_{\lambda}$ is the transit depth expected from the true, wavelength-dependent planet-to-star radius ratio. \textcolor{black}{The numerator within the square root gives the in-transit flux, and the denominator the out-of-transit flux.} \textcolor{black}{In the following sections, we apply the CPAT model to three cases:  1) heterogeneous absorbers, 2) generalized temperature heterogeneities, and 3) cool starspots with parameters fixed to values inferred from long-term photometric monitoring of GJ~1214b.}

\subsubsection{CPAT model for absorber heterogeneities \label{subsec:CPAT-absorber}}

Significant chemical heterogeneities are known to exist for magnetically active hot (B and A-type) main sequence stars that show peculiar chemical abundances \citep[e.g.,][]{Pyper:1969aa, Khokhlova:1985aa}. While no strong correlation was observed between the line profile variations and magnetic field strengths, it has been proposed that the chemical abundance patterns emerge due to anisotropic diffusion of the elements in a strong magnetic field \citep[e.g.,][]{Michaud:1970aa, Urpin:2016aa}. Simultaneous Doppler imaging mapping of chemical heterogeneities and magnetic field geometry argue for highly complex configurations across the stellar disk \citep[e.g.,][]{Piskunov:2002aa}. 
{Similarly, both partially and fully convective mid-M dwarfs have been found to store the bulk of their magnetic flux in small scale components that are non-axisymmetric \citep{reiners2009}.
Field strengths up to $4 \times 10^3$~G have been detected for very active M4.5 dwarfs \citep{johns-krull1996} as well as rapidly rotating mid- to late-M dwarfs \citep{reiners2010, reiners2012}.}
If indeed anisotropic diffusion is the process that leads to chemically heterogeneous stellar photospheres, then it is expected that this effect will only be important for stars with strong magnetic fields ($\sim10^3$~G) and without fully convective atmospheres; however, weaker magnetic fields---not able to produce strong enough chemical heterogeneities to lead to varying absorption line profiles---may lead to low-level heterogeneities that could still potentially influence the optical transmission spectrum of transiting exoplanets. The detailed analysis of this effect is beyond the scope of this paper but may be important for future transit spectroscopy of planets orbiting {stars with strong magnetic fields, including rapidly rotating mid- to late-M dwarfs.}

We use Equation (\ref{eq:CPAT}) to model the effects of an heterogeneous distribution of absorbers in the stellar photosphere on the observed transmission spectrum (the ``CPAT-absorber model''). We utilize PHOENIX model stellar spectra \citep{huss2013} with $T_{eff}=3,300$~K and $\log{g}=5.0$ to generate the $S_{o}$ and $S_{u}$ spectra for our model.  As a proxy for the strength of absorbers, we employ models with a range of metallicities as defined by $[Fe/H]$. Higher metallicity models demonstrate deeper spectral features due to the larger absorber abundances, while lower metallicity models possess relatively muted absorption features. We do not consider alpha-enhanced or depleted models, so $[Fe/H]$ is synonymous with the overall metallicity $Z$. The PHOENIX model grid includes spectra for metallicities of $-4.0 \le [Fe/H] \le -2.0$ in steps of 1.0 and $-2.0 \le [Fe/H] \le +1.0$ in steps of 0.5.  We include each of these models in our analysis.

We employ a Markov chain Monte Carlo approach to find the best-fitting CPAT-absorber model. We conduct an MCMC optimization for each of the exoplanet atmosphere models in the grids of cloudy and hazy models described in Section~\ref{subsec:comparison_model_ts}. At each step in the Markov chain, the measured $R_{p}/R_{s}$ values from the joint dataset are compared to the CPAT-absorber model produced by the combination of the heterogeneous stellar photosphere and the input exoplanet atmosphere. The spectra $S_{o}$ and $S_{u}$ are generated by using the values for $[Fe/H]_{o}$ and $[Fe/H]_{u}$, the metallicities of the occulted and unocculted spectra, respectively, and linearly interpolating between the closest spectra in the PHOENIX model grid. Following the results of Doppler imaging studies of M~dwarfs, we adopt $F=0.032$, which is the mean spot filling factor \citet{barn2015} found for the M4.5~dwarf GJ~791.2A over its rotation period. We fix the metallicity of the occulted spectrum to the best-fit value for the metallicity of GJ~1214, $[Fe/H]_{o}=0.20$ \citep{rojas2012}. The free parameters in the model are the metallicity contrast $\Delta[Fe/H$], which determines the metallicity of the unocculted spectrum relative to that of the occulted spectrum, and the uniform offset applied to the exoplanet's model transmission spectrum, $(R_{p}/R_{s})_{o}$. 
We place a uniform prior on $\Delta[Fe/H]$ to allow the algorithm to fully explore the parameter space, with interval boundaries $[-4.2,+0.8]$ to keep $[Fe/H]_{u}$ within the PHOENIX model range $[-4.0,+1.0]$. {While allowing the metallicity of 3.2\% of the stellar disk to vary will change the mean metallicity of the photosphere, we note that the mean will never vary more than $1\sigma$ from the measured metallicity, $[Fe/H]=0.20\pm0.17$ \citep{rojas2012}.} We place a Gaussian prior on $(R_{p}/R_{s})_{o}$ using the mean and standard deviation of the residuals found by fitting the joint dataset to the exoplanet's model transmission spectrum (Section~\ref{subsec:comparison_model_ts}) and bounded on the interval $[-1,+1]$. We run five chains of $10^{5}$ steps with an additional $10^{4}$ steps discarded as the burn-in. We consider the chains to be well-mixed if {the Gelman-Rubin statistic} $\hat{R} \leq 1.03$ for all parameters.

The results of the CPAT-absorber model fitting for the full grids of cloudy and hazy transmission spectra are illustrated in Figure~\ref{fig:X2_CPAT-abs}.  As can be seen by comparing Figures~\ref{fig:X2_simple} and \ref{fig:X2_CPAT-abs}, the best-fitting parameters for the exoplanet's atmosphere are not substantially changed by incorporating the effect of a heterogeneous stellar photosphere, though the CPAT-absorber models can provide better fits to the data. 
The complete results from the fitting procedure are provided in Tables \ref{table:X2_CPAT-abs_cloud} and \ref{table:X2_CPAT-abs_haze}, for the equilibrium cloud and photochemical haze models respectively.
The values for the free parameters, $\Delta[Fe/H]$ and $(R_{p}/R_{s})_{o}$, that we report there and quote below are the median and 68\% confidence intervals from the MCMC optimization procedure. 
In the case of equilibrium clouds, we find the best-fitting model to have a very high metallicity and thick clouds in the exoplanet's atmosphere and {a relatively low metallicity in the unocculted region of the star's photosphere}
($1000\times$~solar, $f_{sed}=0.1$, $\Delta[Fe/H]=-1.58^{+0.28}_{-0.37}$, ${(R_{p}/R_{s})_{o}=8,594^{+28}_{-26}}$~{ppm,} $\chi^{2}=54.7$, $DOF=36$, $\chi^{2}_{red}=1.52$).
Considering photochemical haze models, we find the best-fitting model to include the same mode particle size and haze-forming efficiency as when considering only the exoplanetary contribution to the transmission spectrum (Section~\ref{subsec:comparison_model_ts}) and a low metallicity for the unocculted region similar to that found for the CPAT-absorber cloud model 
($r=0.3~\micron$, $f_{haze}=10\%$, $\Delta[Fe/H]=-1.31^{+0.35}_{-0.39}$, ${(R_{p}/R_{s})_{o}=-5,758^{+27}_{-26}}$~{ppm,} $\chi^{2}=53.3$, $DOF=36$, $\chi^{2}_{red}=1.48$). Figure~\ref{fig:posteriors_CPAT-abs} shows the posterior distributions for the free parameters in each of the best-fitting cloud and haze models.
{A model using a constant, achromatic value of $R_{p}/R_{s}$ for the exoplanet's transmission spectrum \textbf{together with a CPAT-absorber model for the star} provides a fit to the data comparable to that of the best-fitting CPAT-absorber cloud and haze models 
($\Delta[Fe/H]=-1.02^{+0.33}_{-0.24}$, $R_{p}/R_{s}=0.11604 \pm 3 \times 10^{-5}$, $\chi^{2}=55.8$, $DOF=36$, $\chi^{2}_{red}=1.55$).}

Figure \ref{fig:ts_modelfits_CPAT-abs} shows the best-fitting equilibrium cloud and photochemical haze CPAT-absorber models. Additionally shown is {the constant-${R_{p}/R_{s}}$} CPAT-absorber model, which illustrates the effect of {a} heterogenous distribution of absorbers in the stellar photosphere on an otherwise flat transmission spectrum. It shows that variations in the strength of absorbers across the stellar disk can produce large deviations from a flat spectrum in the optical while simultaneously preserving a flat transmission spectrum in the near-infrared.  

{For both the cloud and haze cases, the best fits to the data using the CPAT-absorber model require 3.2\% of the unocculted stellar disk to possess weaker absorption features than the transit chord.} In effect, the unocculted stellar disk {is} brighter in {optical} absorption bands than the region of the photosphere typified by the transit chord, which decreases the observed $R_{p}/R_{s}$ within those absorption bands. {This allows the CPAT-absorber model to provide better fits to the data than the flat line model or models for atmospheric transmission alone considered in Section~\ref{subsec:comparison_model_ts} but still does not fully explain the observed spectrum.}

\subsubsection{CPAT model for temperature heterogeneities \label{subsec:CPAT-temperature}}

We also investigate the contribution of stellar temperature heterogeneities to the observed transmission spectrum using the CPAT model (the ``CPAT-temperature model''). We employ {PHOENIX models} with {a surface gravity} ($\log{g}=5.0$) and {a metallicity} ($[Fe/H]=0.0$){, the closest PHOENIX grid values to those identified for GJ~1214 by \citet{rojas2012},} and temperatures in the range $2,700~\textrm{K} \le T_{eff} \le 4,700~\textrm{K}$. We conduct the MCMC \textcolor{black}{optimization} procedure as described in Section \ref{subsec:CPAT-absorber}, {linearly} interpolating {in temperature space} within the PHOENIX model grid to generate the $S_{o}$ and $S_{u}$ spectra based on the effective temperatures of the occulted and unocculted regions ($T_{o}$ and $T_{u}$, respectively) at each step. We fix $T_{o}=3,252$~K using the effective temperature of GJ~1214 \citep{angl2013} and $F=0.032$ following the results of \citet{barn2015}. The free parameters in the model are the temperature contrast $\Delta T$, which determines $T_{u}$ relative to $T_{o}$, and $(R_{p}/R_{s})_{o}$. We place a Gaussian prior on $\Delta T$ centered on 0~K with an uncertainty equal to 10\% of the $T_{eff}$ of GJ~1214 to thoroughly explore the parameter space and truncate it on the range $[-552~\textrm{K}, +1,448~\textrm{K}]$ to keep $T_{u}$ within the PHOENIX model range we consider $[2,700~\textrm{K}, \mathbf{4,700}~\textrm{K}]$. 
As with the CPAT-absorber model, we conduct an MCMC optimization for each of the exoplanet atmosphere models in the grids of cloudy and hazy models described in Section~\ref{subsec:comparison_model_ts}.

Figure~\ref{fig:X2_CPAT-temp} illustrates the goodness-of-fit for the full grids of cloudy and hazy transmission spectra modulated by the CPAT-temperature model. As with the CPAT-absorber model, the best-fitting parameters for the exoplanet's atmosphere are similar to those found when considering the contribution of the exoplanet atmosphere alone to the transmission spectrum (Figure~\ref{fig:X2_simple}). However, allowing for temperature heterogeneities provides better fits to the data than either the exoplanet-alone or CPAT-absorber models. The complete results for the CPAT-temperature model fitting for the equilibrium cloud and photochemical haze model grids are provided in Tables~\ref{table:X2_CPAT-temp_cloud} and \ref{table:X2_CPAT-temp_haze}, respectively.  The best-fitting cloud model includes an exoplanet atmosphere with the highest metallicity and thickest clouds of the models we considered and a temperature contrast of $\Delta T \simeq + 0.1 T_{eff}$ for the unocculted region of the star's photosphere 
($1000\times$~solar, $f_{sed}=0.01$, $\Delta T=354^{+46}_{-47}$~K, {${(R_{p}/R_{s})_{o}={9,054}^{+110}_{-94}}$~{ppm}, $\chi^{2}=45.2$, $DOF=36$, $\chi^{2}_{red}=1.25$). 
The best-fitting haze model includes a smaller mode particle size and the same haze-forming efficiency as the best-fitting exoplanet-only model and $\Delta T \simeq + 0.1 T_{eff}$ as well 
($r=0.1~\micron$, $f_{haze}=10\%$, $\Delta T=354^{+46}_{-46}$~K, {${(R_{p}/R_{s})_{o}={-3,064}^{+101}_{-102}}$~{ppm}, $\chi^{2}=40.5$, $DOF=36$, $\chi^{2}_{red}=1.13$). 
This model is the best-fitting of all those we considered in this analysis.  The posterior distributions for the free parameters in each of the best-fitting cloud and haze CPAT-temperature models are shown in Figure~\ref{fig:posteriors_CPAT-temp}. {The $\Delta T$ and $(R_{p}/R_{s})_{o}$ parameters are positively correlated due to the fact that larger temperature contrasts depress model $R_{p}/R_{s}$ values at all wavelengths and thus require larger offsets to bring models in line with observations.}
{A model using a constant $R_{p}/R_{s}$ value for the exoplanet's transmission spectrum \textbf{together with a CPAT-temperature model for the star} provides a fit to the data comparable to that of the best-fitting CPAT-temperature cloud model but not as good as that of the best-fitting CPAT-temperature haze model 
($\Delta T=336^{+54}_{-45}$~K, $R_{p}/R_{s}=0.11680 \pm 1 \times 10^{-4}$, $\chi^{2}=45.6$, $DOF=36$, $\chi^{2}_{red}=1.27$).}

The transmission spectra for the best-fitting equilibrium cloud and photochemical haze CPAT-temperature models are shown in Figure \ref{fig:ts_modelfits_CPAT-temp}. The constant $R_{p}/R_{s}$ model, shown as a dashed black line, illustrates the effect of a temperature heterogeneity in the stellar photosphere on an otherwise flat transmission spectrum. Its similarity in the optical to the cloud and haze models owes to the fact that the best-fitting exoplanet transmission spectra are essentially flat in the optical and the observed variations are due to features imprinted by the heterogeneous stellar photosphere.
In each of these cases, a region of the unocculted stellar disk is brighter than that occulted by the transiting planet, which effectively decreases the observed $R_{p}/R_{s}$ during the transit. The effect is chromatic, producing the largest change in $R_{p}/R_{s}$ in the optical, where the difference in emergent flux between the stellar spectral models is most pronounced.

{Cool, unocculted spots, on the other hand, would increase the observed $R_{p}/R_{s}$ values. W}e use the CPAT-temperature model to investigate how cool, unocculted starspots {reported in the literature} could affect the transmission spectrum. Long-term monitoring shows that GJ~1214 demonstrates a $1\%$ peak-to-peak variability in the MEarth \citep{nutz2008} bandpass ($715~\textrm{nm} < \lambda < 1,000~\textrm{nm}$) on a timescale that is an integer multiple of 53 days \citep{bert2011}, which has been attributed to rotational modulation of cool starspots \citep{char2009, bert2011, frai2013}. We calculate the spot-covering fraction implied by the variability using PHOENIX model spectra with $\log{g}=5.0$ and $[Fe/H]=0.0$. Assuming the spots are 10\% cooler than the unspotted surface, following results from Doppler imaging of M dwarfs \citep{barn2015}, we utilize models with $T=3,252$~K and $T=2,927$~K for the unspotted and spotted surfaces, respectively, by linearly interpolating {in temperature space} between models in the PHOENIX grid. Integrating over the MEarth bandpass\footnote{\url{http://newton.cx/~peter/wp/wp-content/uploads/2014/08/mearth-bandpass.html}}, we find that a spot-covering fraction $F=0.03$ of the stellar disk can reproduce the reported variability \textcolor{black}{in the MEarth bandpass}. The effect of such a spot configuration on the observed transmission spectrum is shown in gray on Figure \ref{fig:ts_modelfits_CPAT-temp}. In the near-infrared, the scale of the effect is smaller than the reported measurement uncertainties. The most pronounced change is present in the optical, where the observed transmission spectrum is increased as much as ${\Delta(R_{p}/R_{s})=8 \times 10^{-4}}$ above the near-infrared spectrum. {The optical transmission spectrum would decrease by a similar amount if these starspots were instead occulted by the transiting exoplanet \citep{pont2013}. However, the observed decrease in the optical transmission spectrum is $\sim 3 \times$~larger than what would be caused by these spots.  Additionally, no brightening events due to occulted starspots are evident in our data (Figure~\ref{fig:wl}).}

In summary, we find that cool starspots cannot account for the decreased optical $R_{p}/R_{s}$ values, though unocculted bright regions of the photosphere with $\Delta T \simeq + 0.1 T_{eff}$ can decrease a flat optical transmission spectrum to the observed values.

\subsubsection{Physical interpretation of CPAT model  \label{sec:interpretation}}

The results of our modeling efforts indicate that the deviations from a flat optical transmission spectrum for GJ~1214b could be introduced by heterogeneities in the stellar photosphere, either in terms of temperature or the distribution of absorbers (Figures~\ref{fig:ts_modelfits_CPAT-abs} and \ref{fig:ts_modelfits_CPAT-temp}). 
However, the near-infrared transmission spectrum is largely unaffected by the stellar photosphere. The combination of the optical and near-infrared data, then, allows us to probe the stellar and planetary contributions to the transmission spectrum simultaneously.

The optical spectrum of a M4.5 dwarf like GJ~1214 is largely driven by opacity from TiO molecular bands \citep{morg1943}. Using the CPAT-absorber model framework, we find that an unocculted region of the stellar photosphere with relatively weak absorption in these bands could imprint spectral features on an otherwise flat exoplanetary transmission spectrum.  However, the best-fitting CPAT-absorber model requires a metallicity contrast of $\Delta[Fe/H]=-1.31$, corresponding {to} a depletion in the abundance of absorbers by a factor of 20 in the unocculted region.  Efforts to measure a latitudinal dependence of the solar spectrum have found elemental abundances to be within 0.005~dex across latitudes \citep{kise2011}. Moreover, as
stars with masses less than 0.35~$M_{\odot}$ (corresponding to spectral types M3.5 and later) are fully convective \citep{chab1997}, a viable mechanism for maintaining significant sustained abundance differences in the stellar photosphere of a M4.5 dwarf like GJ~1214 is not immediately apparent.

Temperature heterogeneities, however, are known to exist in stellar photospheres.
We find a temperature heterogeneity could produce the observed {discrepancy in $R_{p}/R_{s}$ between the optical and near-infrared} if $\sim3.2\%$ of the unocculted stellar disk is $\sim350$~K hotter than the remaining photosphere.  GJ~1214 is known to host cool starspots from starspot-crossing events detected in transit light curves \citep{cart2011, krei2014}. Mid- to late-M dwarfs have also been found to have abundant polar spots from Doppler imaging. \citet{barn2015} found the M4.5 dwarf GJ~791.2A to have a mean spot coverage of 3.2\% and a maximum spot coverage of 82.3\% during its rotation, assuming spots are 300~K cooler than the photosphere. On the Sun, spots are accompanied by bright plage, which include faculae with temperature contrasts of 300--500~K with their surroundings \citep{topk1997}. Brightening from these faculae overpowers spot darkening, causing total solar irradiance to increase during solar maxima \citep{frohlich1998, meunier2010}. Faculae cover roughly 0.36\% of the sky-projected solar disk during periods of low solar activity and 3\% during high activity \citep{shap2014}. Additionally, faculae are much more common than spots on the Sun, covering $100\times$~more disk-area during periods of low activity, and $10\times$~more area during periods of high activity \citep{shap2014}.

{Like the Sun, o}ld, slowly rotating FGK stars are known to have activity cycles dominated by faculae, unlike their younger counterparts, which are spot-dominated \citep{radick1983, lockwood2007}. With an age of 3--10~Gyr \citep{char2009}, the photometric variability of GJ~1214 may be faculae-dominated as well. {O}ur data are most consistent with unocculted faculae in the photosphere of GJ~1214 producing the observed {offset} in the optical transmission spectrum of GJ~1214b.

\subsubsection{{Observational considerations of heterogeneous stellar photospheres}}

The results of our CPAT-absorber and CPAT-temperature modeling efforts demonstrate that, in principle, a heterogeneous stellar photosphere can provide a transmission spectroscopy signal larger than that introduced by the exoplanetary atmosphere in the optical, adding another degeneracy to the modeling and interpretation of exoplanet spectra.
At longer wavelengths, however, the effect is less pronounced \textcolor{black}{for a star with the $T_{eff}$ of GJ~1214}, and differences between the atmospheric models are more apparent.  We model the effect out to the maximum wavelength of the PHOENIX stellar models ($5.5~\micron$), and find the largest differences between transmission spectra for cloudy and hazy atmospheres at 2--5~$\micron$ for both CPAT-absorber and CPAT-temperature cases. This would be a promising region to target in transmission spectroscopy studies with the James Webb Space Telescope (JWST) in order to distinguish between cloudy and hazy atmospheres for this planet. Optical spectra like that presented here can complement JWST investigations of GJ~1214 and other targets by constraining the contribution of heterogeneous stellar photospheres to observed transmission spectra.

This analysis suggests that facular brightening may contribute more to transmission spectra of exoplanets than has been previously recognized.
If common for M dwarfs, this photospheric heterogeneity could complicate optical transmission spectroscopy studies of exoplanets around small stars, such as GJ~1132b \citep{bert2015} and TRAPPIST-1b, c, and d \citep{gillon2016}, including searches for Rayleigh scattering. While unocculted starspots can mimic the transmission signature of scattering in an exoplanetary atmosphere \citep[e.g.,][]{McCullough:2014aa}, facular brightening has the potential to mask a scattering signature by reducing the Rayleigh scattering slope in a transmission spectrum. However, the method we provide here can be used to take unocculted faculae into account if high SNR transmission spectra are obtained. Additionally, the same exoplanets would provide a unique opportunity to gain spatial information about M dwarf surfaces if the exoplanetary and photospheric contributions can be uniquely identified. The Transiting Exoplanet Survey Satellite (TESS), which will utilize a long-pass filter from 0.6--1.0~$\micron$, will measure transit depths for potentially hundreds of planets around M dwarfs; comparing the TESS transit depths to follow-up measurements in the near-infrared could be one way of probing for spectral features introduced by a heterogeneous stellar photosphere.  Along a different avenue of research, future work can investigate the range of stellar temperatures and metallicities for which this effect will be important.

\section{SUMMARY \label{sec:summary}}

We have presented an optical transmission spectrum of the sub-Neptune GJ~1214b measured during three transits with Magellan/IMACS. The spectrum, which covers 4,500 to 9,260~\AA~in 14 bins with a mean value of $R_{p}/R_{s}=0.1146\pm2\times10^{-4}$ and mean uncertainty of $\sigma(R_{p}/R_{s})=8.7\times10^{-4}$, is offset below the near-infrared $R_{p}/R_{s}$ values previously reported and cannot be reproduced by cloud/haze models for this exoplanet. {We summarize below the k}ey points from this study:  
\begin{enumerate}  
\item We find consistent spectra from three different transits taken with two different instrument configurations and reduced with different approaches, resulting in one of the most robust ground-based transmission spectra of a sub-Jovian exoplanet.
\item We find that the optical transit depth is shallower than that measured in the near-infrared{. The data hint at more variation in $R_{p}/R_{s}$ at optical wavelengths than has been observed in the near-infrared.}
\item We describe a new model, CPAT, {that can be used to evaluate the effect of heterogeneous stellar photospheres in the interpretation of exoplanet transmission spectra.}
\item {We use the CPAT model in the case of GJ~1214b.} We find that {the} data are not consistent with a perfectly homogeneous stellar photosphere or a photosphere that is compositionally heterogeneous but isothermal. 
\item {The} data are {fit best by a model with thick haze ($r=0.1~\micron$, $f_{haze}=10\%$) in the exoplanet's atmosphere and hotter photospheric features covering 3.2\% of the unocculted stellar disk with a temperature contrast $\Delta T=354^{+46}_{-46}$~K.} The parameters of these features are consistent with {those of solar faculae.}
\item Our results highlight the importance of heterogeneous stellar photospheres for the correct interpretation of optical transmission spectra of transiting planets {and show} that transiting planets may be used as probes of stellar photospheric features.
\end{enumerate}

\acknowledgments{We thank the anonymous referee for their suggestions and comments, which helped greatly to improve the manuscript. This paper includes data gathered with the 6.5 meter Magellan Telescopes located at Las Campanas Observatory, Chile. We thank the operations staff at Las Campanas Observatory for their support with the observations. We thank the University of Arizona, Harvard, Carnegie, Massachusetts Institute of Technology, and Chilean National TACs for allocating the telescope time for \textit{ACCESS}. B.R. acknowledges support from the National Science Foundation Graduate Research Fellowship Program under Grant No. DGE-1143953. N.E. is supported by CONICYT-PCHA/Doctorado Nacional. A.J. acknowledges support from FONDECYT project 1130857, the Ministry of Economy, Development, and Tourism's Millennium Science Initiative through grant IC120009, awarded to The Millennium Institute of Astrophysics, MAS, and BASAL CATA PFB�06. The results reported herein benefited from collaborations and/or information exchange within NASA's Nexus for Exoplanet System Science (NExSS) research coordination network sponsored by NASA's Science Mission Directorate.
}

\bibliography{gj1214b}

\clearpage

\begin{figure}
    \plotone{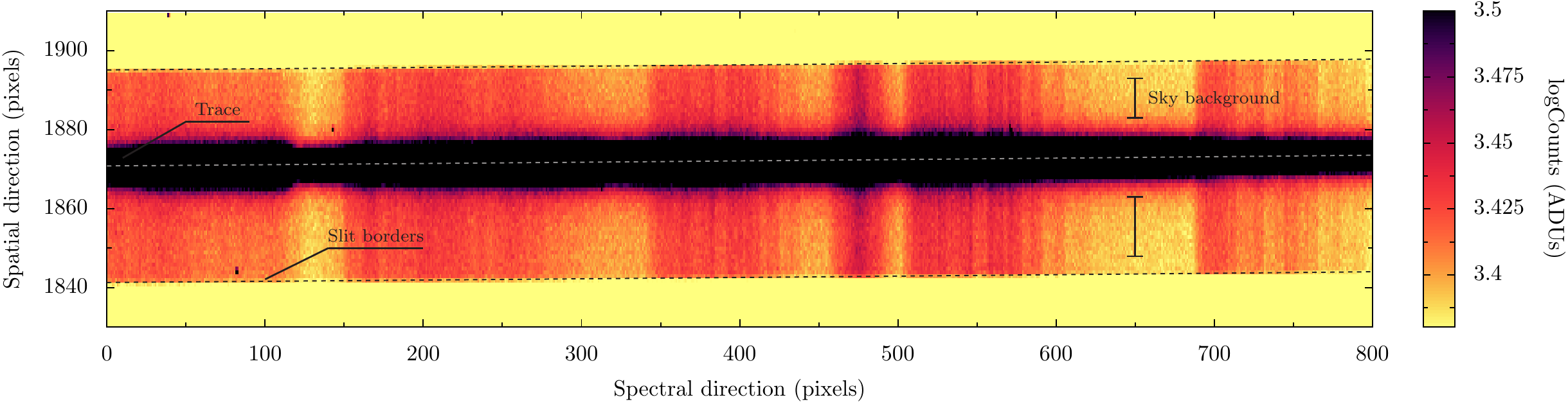}
    \caption{{\bf Example of data showing a portion of the slit containing the spectrum of GJ~1214.} The spectral trace and the slit borders are shown as dashed lines. {A typical region used to measure the sky background is shown as vertical bars.} This sub-image is taken from the Transit~1 dataset. \label{fig:data_example}}
\end{figure}

\begin{figure}
    \plotone{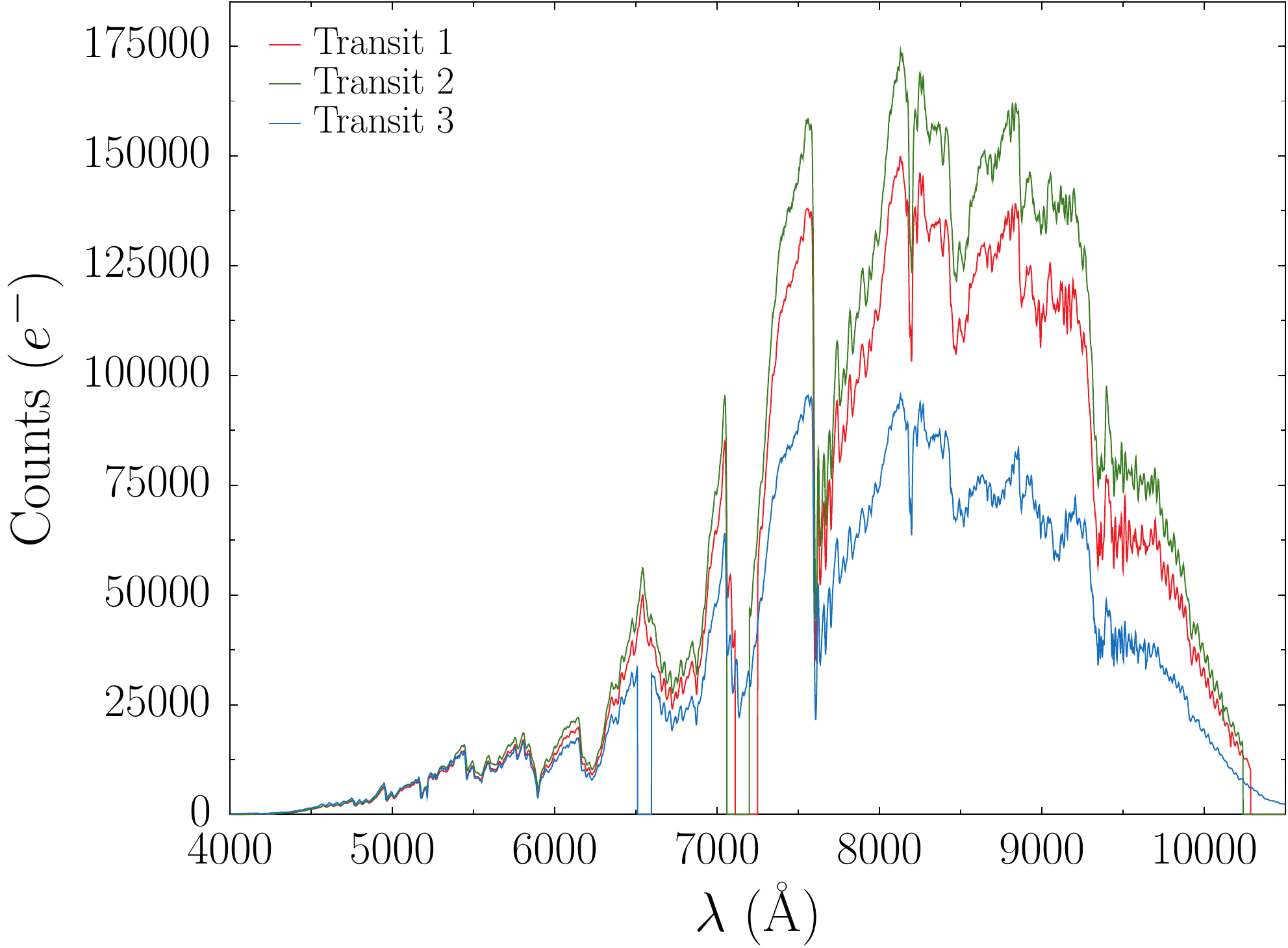}
    \caption{{\bf Median spectra of GJ~1214 from Transits 1 ({\em red}), 2 ({\em green}), and 3 ({\em blue})}. \textcolor{black}{We observed Transits 1 and 2 using the IMACS f/4 camera, and Transit~3 with the f/2 camera.} The exposure times we utilized account for the difference in maximum counts between Transits 1 and 2. The overall shape of the spectrum from Transit~3 differs from that of Transits 1 and 2 due to the transmission profile of the IMACS f/2 camera. Vertical lines indicate chip gaps.
\label{fig:spectra}}
\end{figure}

\begin{figure}
    \plotone{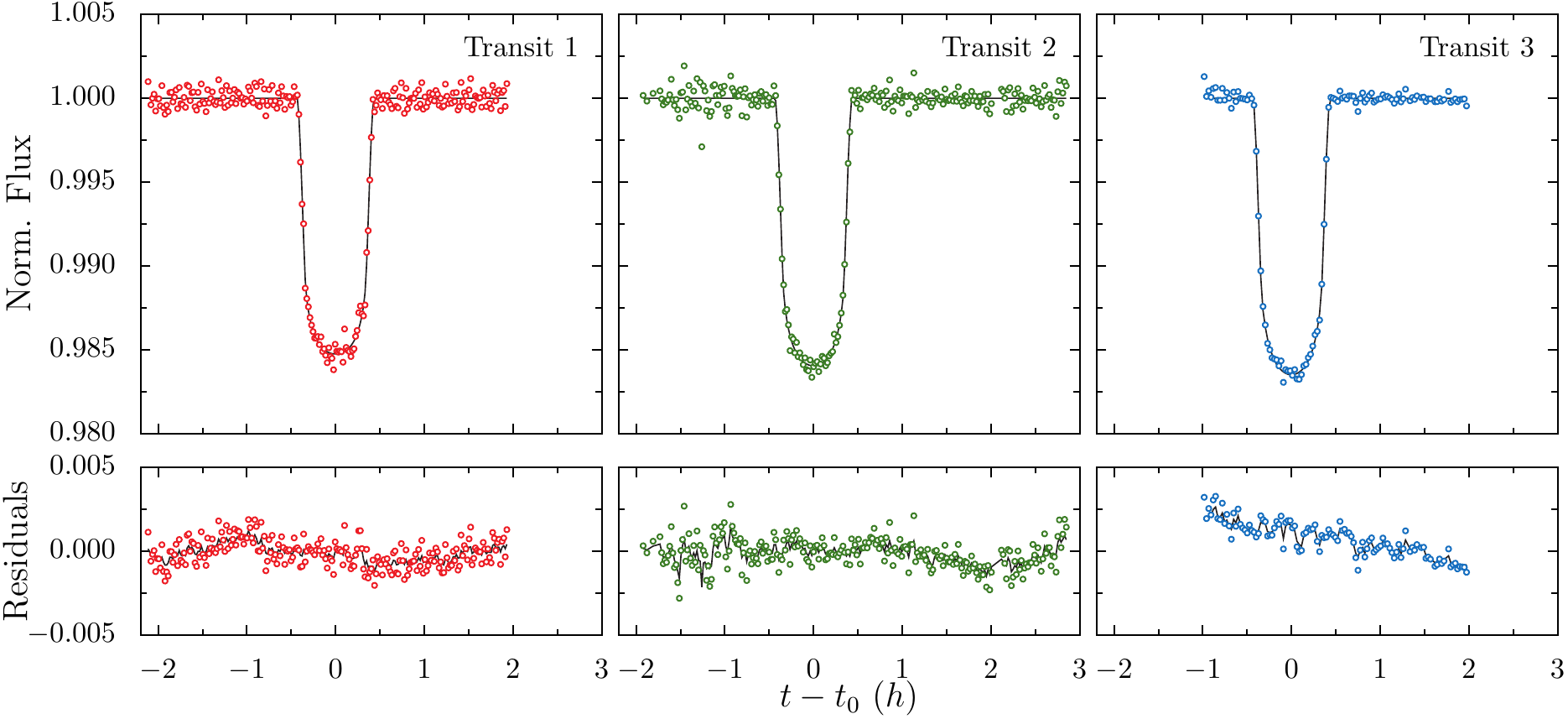}
    \caption{{\bf White light curves ({\em top}) and residuals ({\em bottom}) from Transits 1 ({\em left}), 2 ({\em center}), and 3 ({\em right})}. The top panels show the normalized flux measurements of GJ~1214 \textbf{after subtracting the correlated noise component (colored points)} and the best-fitting transit model for each night \textbf{(black lines)}. The bottom panels show the residuals between the normalized flux measurements and best-fitting transit models {(points)} along with the correlated noise components identified by the wavelet analysis (black lines).
\label{fig:wl}}
\end{figure}

\begin{figure}
    \plotone{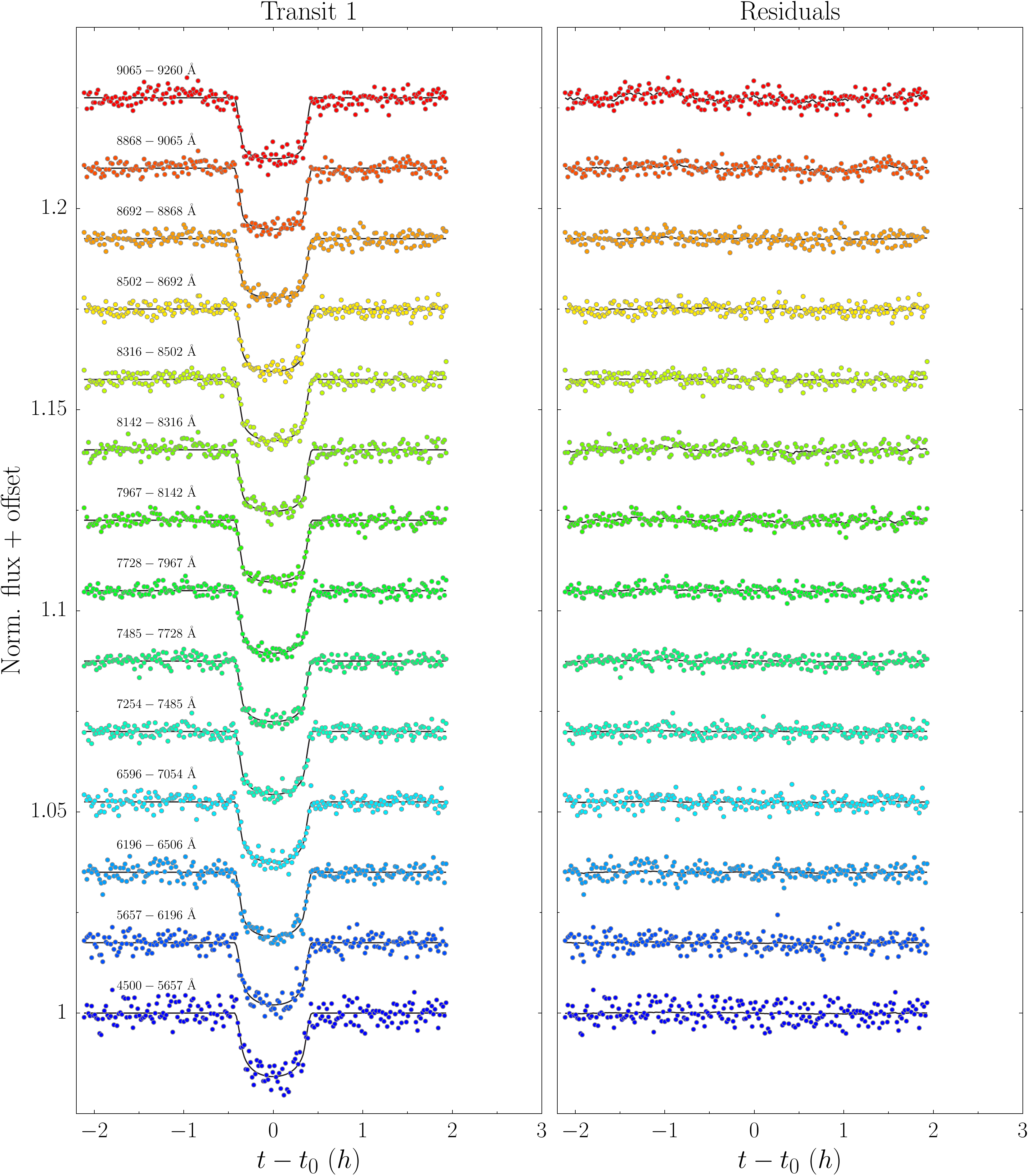}
    \caption{{\bf Detrended light curves and residuals from Transit~1 using the polynomial-based detrending procedure}. The best fitting transit models are plotted as black lines in the left panel. Black lines in the right panel show the correlated noise components identified by the wavelet analysis. Data from different wavelength bins are offset for clarity. Correlated noise levels are {smaller than the Poisson noise} for all wavelength bins.
\label{fig:t1poly}}
\end{figure}

\begin{figure}
    \plotone{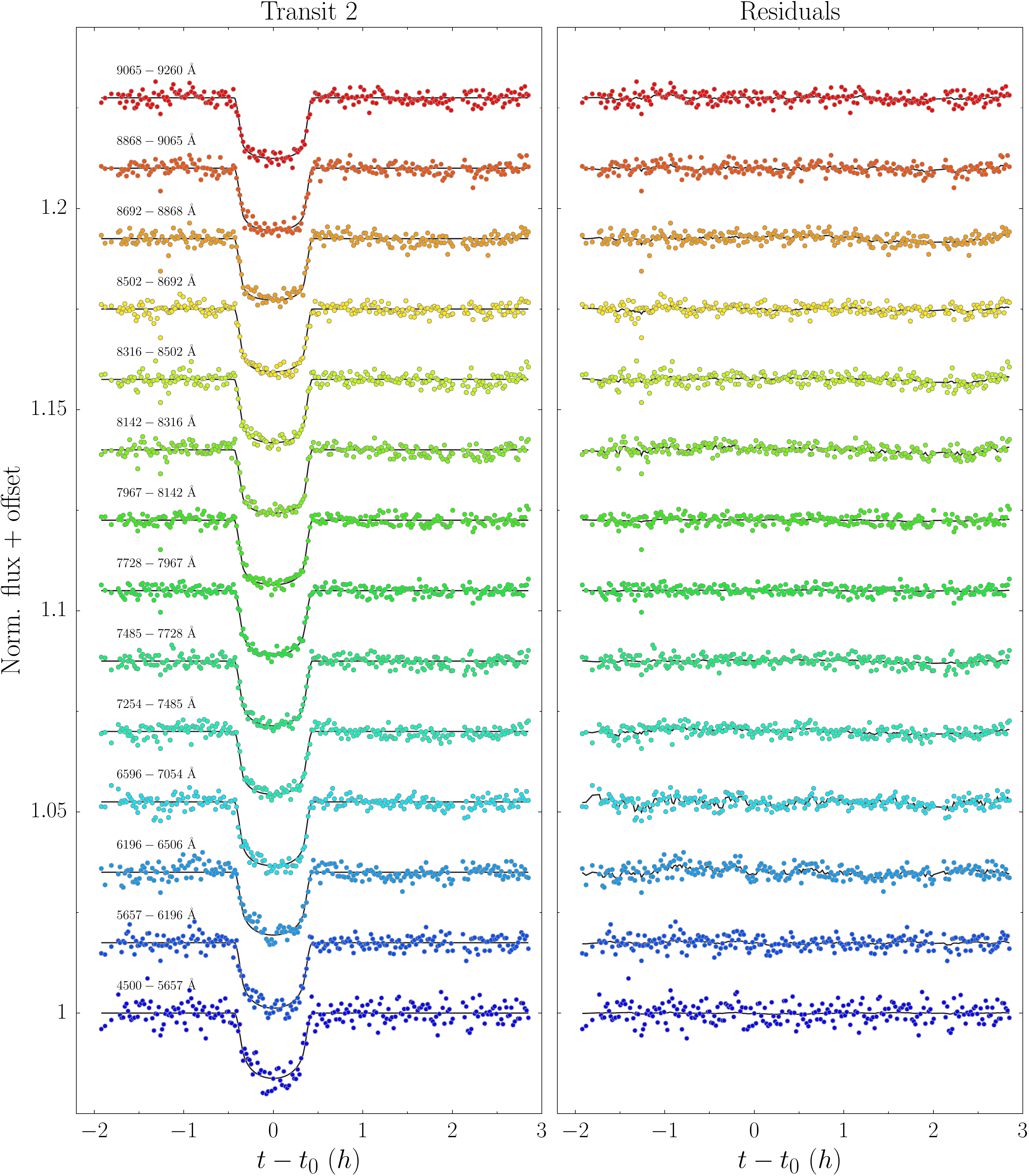}
    \caption{{\bf Detrended light curves and residuals from Transit~2 using the polynomial-based detrending procedure}. The figure components are the same as those for {\em Fig. \ref{fig:t1poly}}.
    \label{fig:t2poly}}
\end{figure}

\begin{figure}
    \plotone{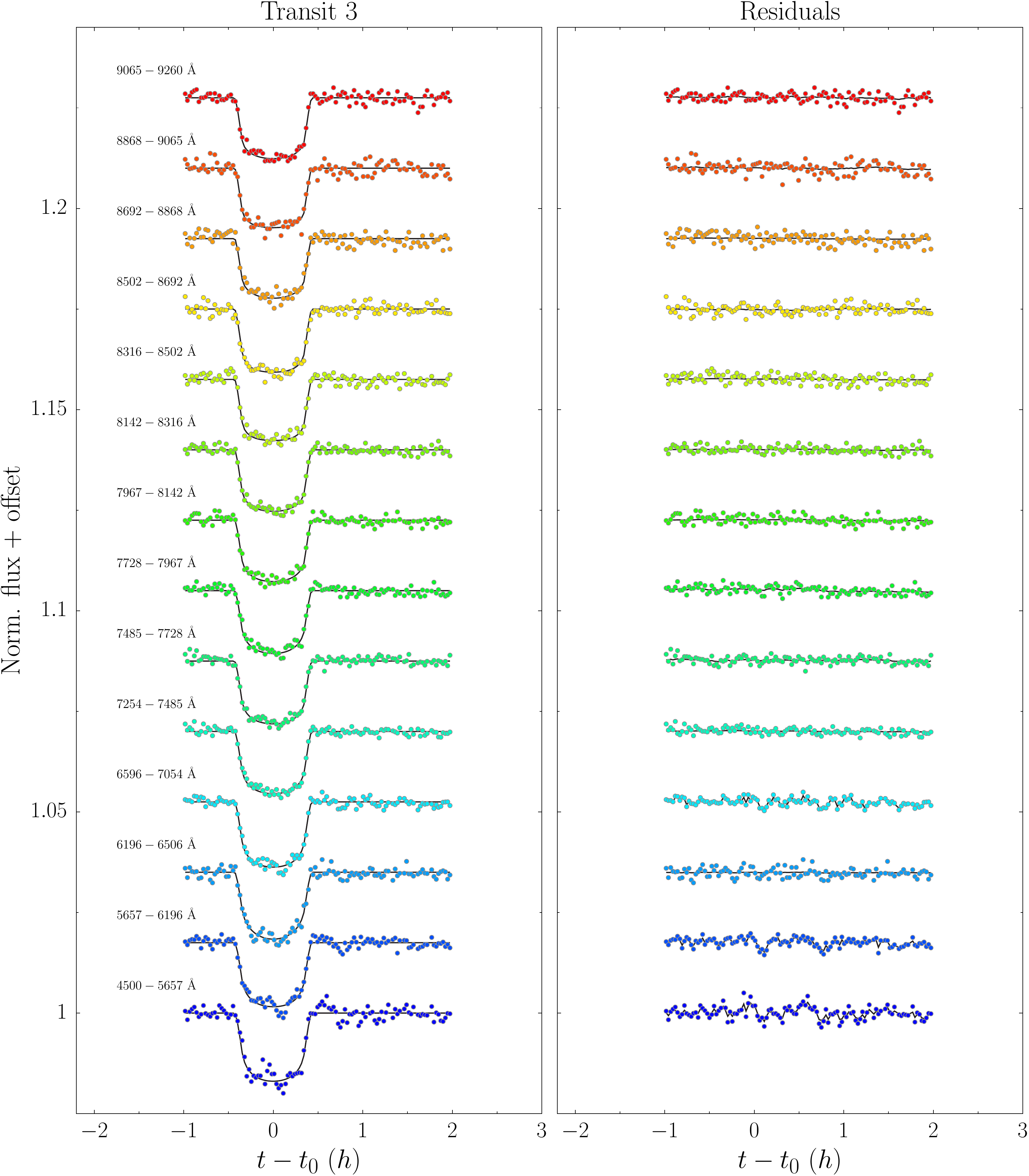}
    \caption{{\bf Detrended light curves and residuals from Transit~3 using the polynomial-based detrending procedure}. The figure components are the same as those for {\em Fig. \ref{fig:t1poly}}. With longer exposures and shorter duty cycles, photon noise levels are lower for the Transit~3 dataset, which utilized the IMACS f/2 camera. Time-correlated systematics are strongest for the bluest wavelength bins.
    \label{fig:t3poly}}
\end{figure}

\begin{figure}
    \plotone{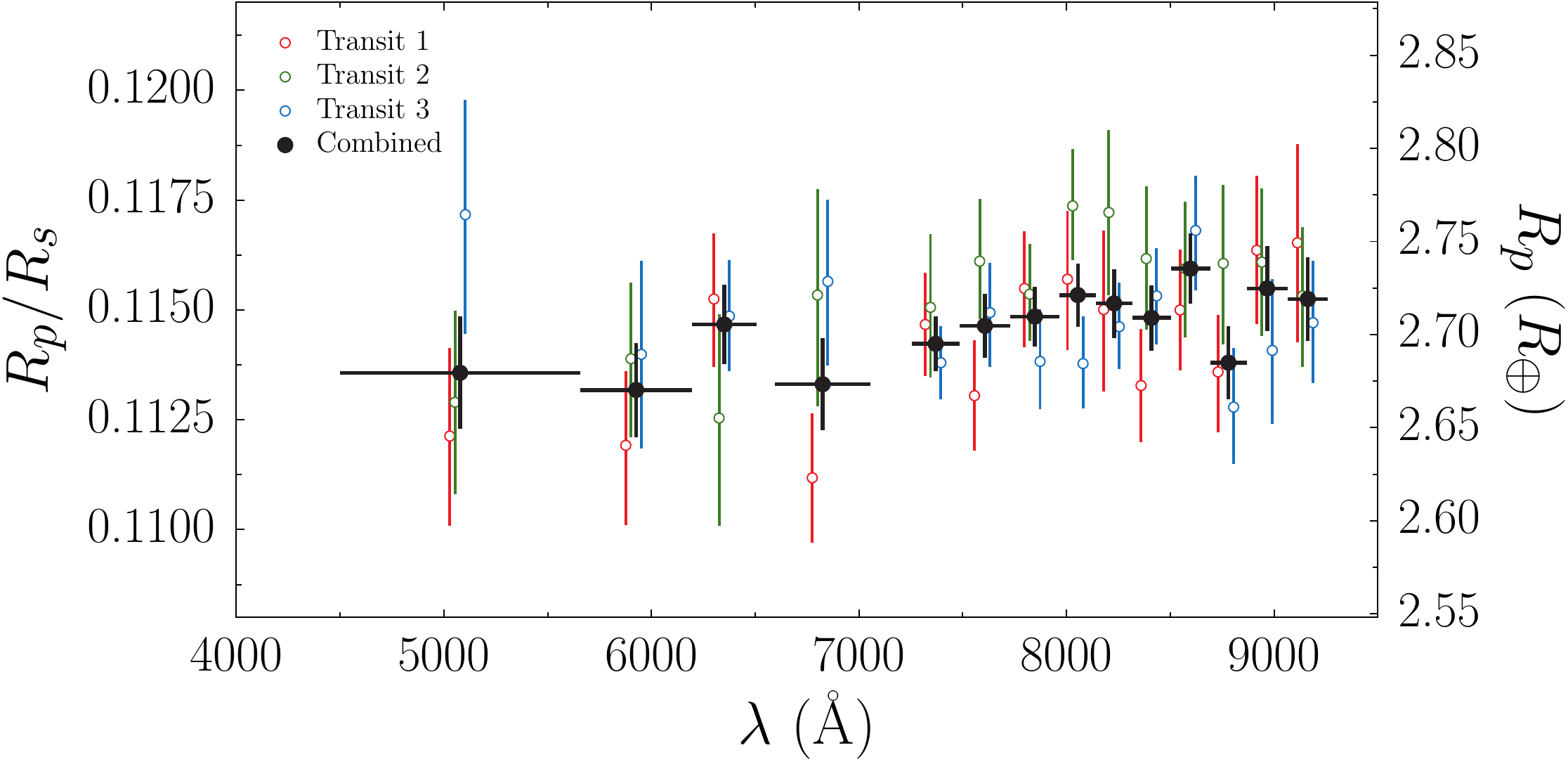}
    \caption{{\bf Transmission spectrum of GJ~1214b from Magellan/IMACS}. Transits 1, 2, and 3 are shown in red, green, and blue, respectively, with horizontal offsets for clarity. The final, combined spectrum (see Section~\ref{subsec:combining_results}) is shown in black. The horizontal error bars on the final spectrum indicate the wavelength bins utilized for all transits.
    \label{fig:tspoly}}
\end{figure}

\begin{figure}
   \epsscale{1}
    \plotone{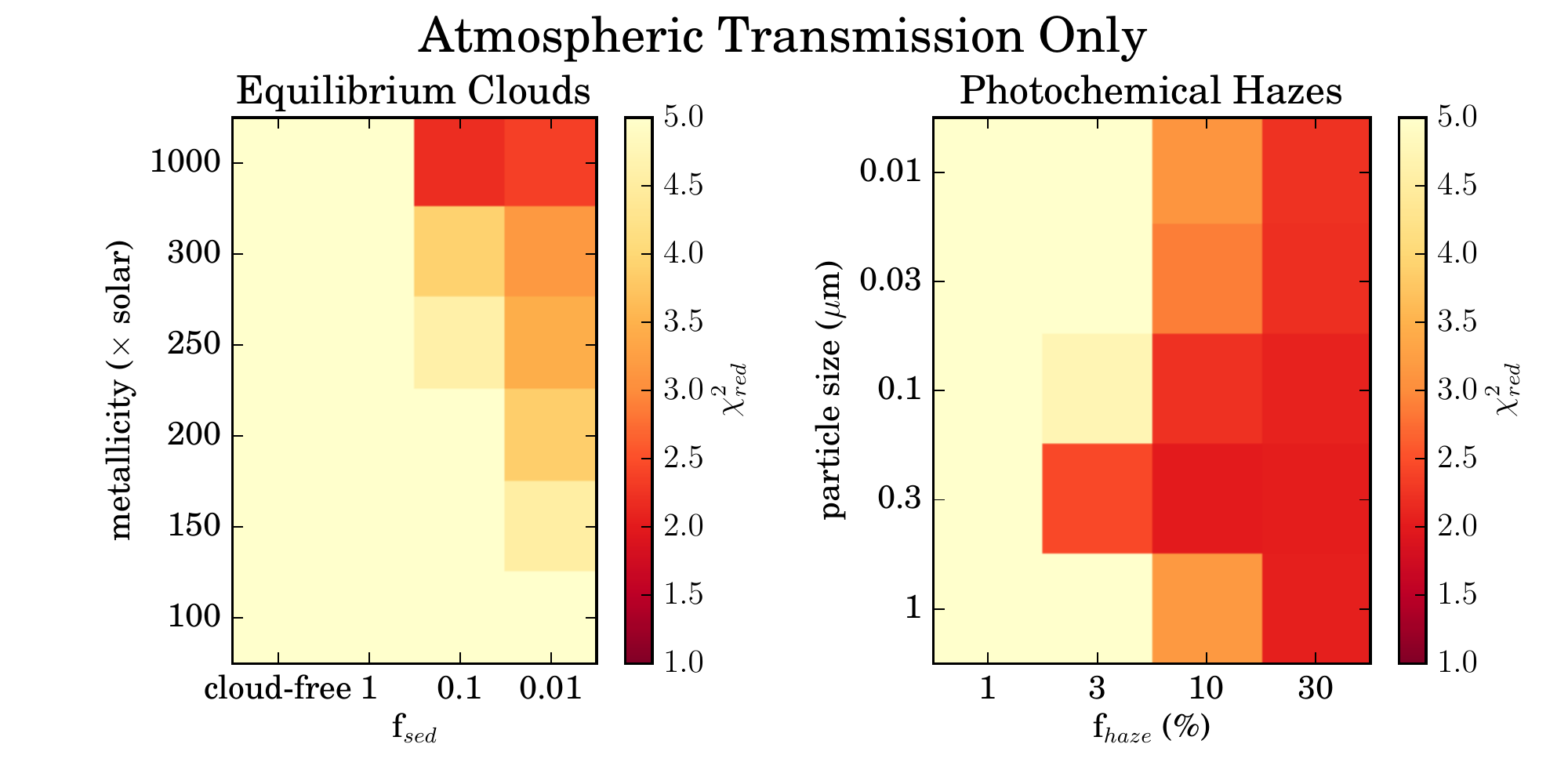}
    \caption{{\bf Reduced chi-squared maps for model fits considering only transmission through the exoplanet's limb.}
    We find the best fits to the joint dataset for cloudy models with high metallicity ($1000\times$~solar) and thick clouds ($f_{sed}\sim0.01$--0.1) and hazy models with relatively large particles (0.1--1$~\micron$) and high haze-forming efficiencies (10--30\%). The full results for the cloudy and hazy model grids \citep{morl2015} are provided in Tables~\ref{table:X2_simple_cloud} and \ref{table:X2_simple_haze}, respectively.
    \label{fig:X2_simple}}
\end{figure}

\begin{figure}
   \epsscale{0.9}
    \plotone{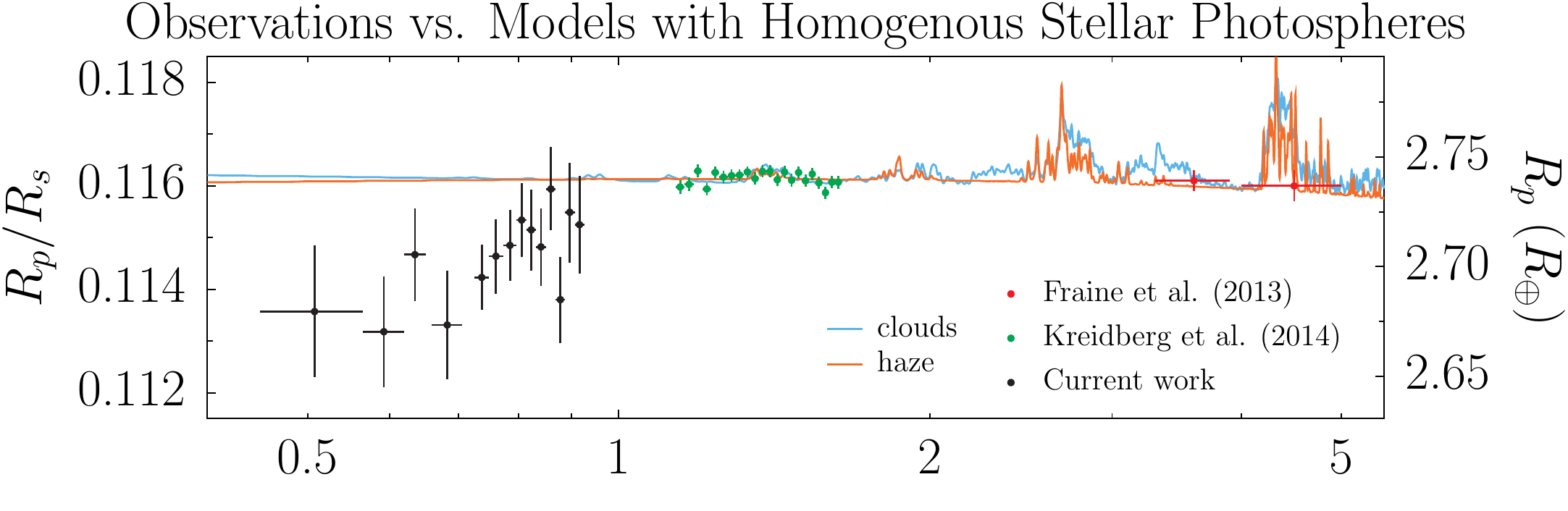}
    \plotone{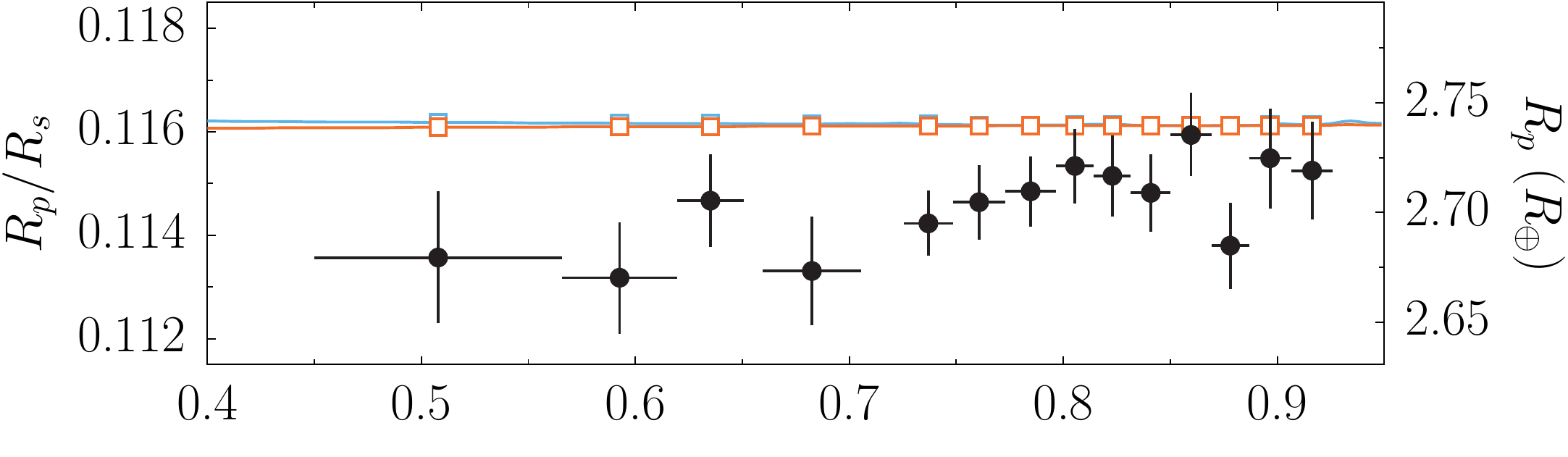}
    \plotone{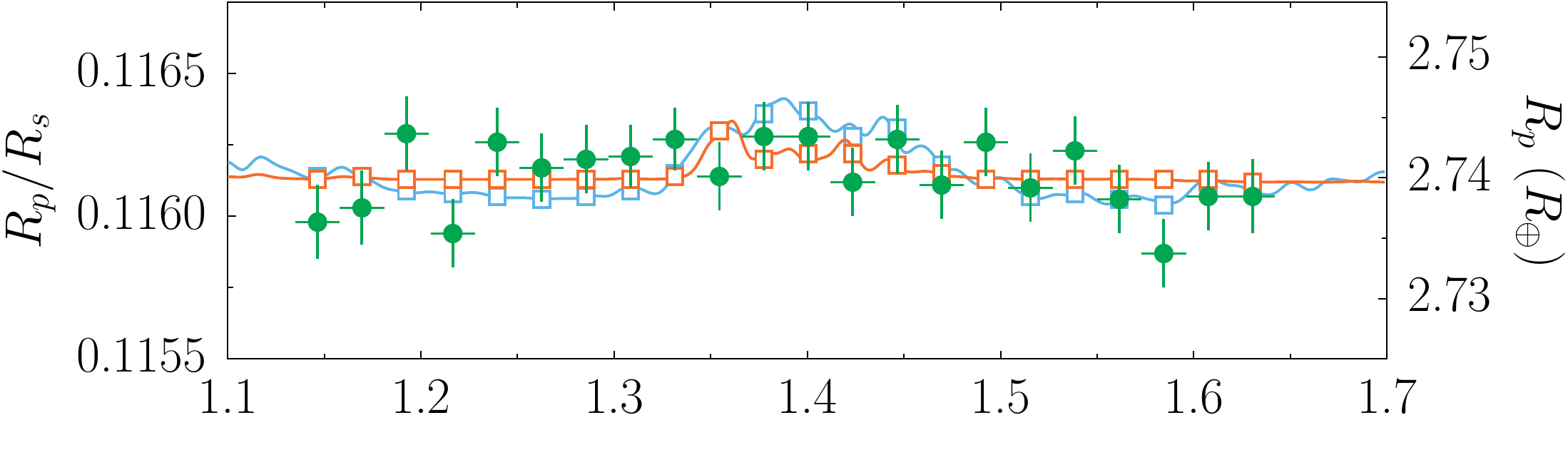}
    \plotone{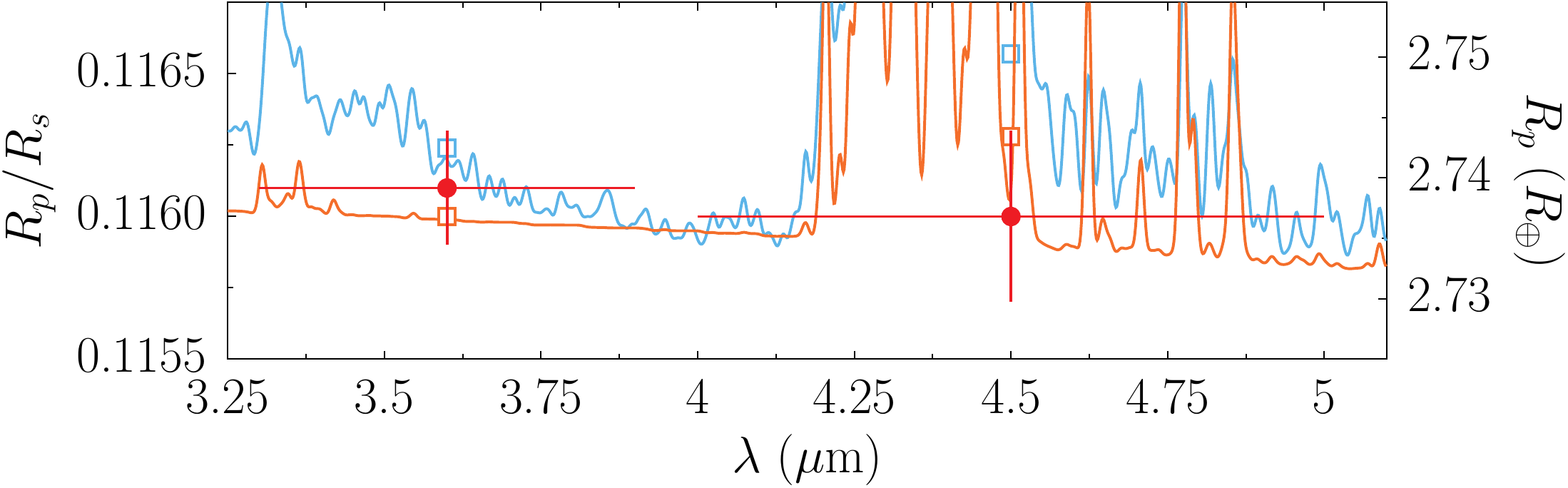}
    \caption{{\bf Best-fitting model spectra only considering atmospheric transmission through the exoplanet's limb.} Models assuming homogeneous stellar photospheres cannot reproduce both the near-infrared and optical measurements. From top to bottom, the panels show the full wavelength range of the best-fit cloudy (1000$\times$~solar, $f_{sed}=0.1$) and hazy ($K_{zz}=10$, $r=0.3~\micron$, $f_{haze}=10\%$) models, and close-up views of the regions around the Magellan (current work), HST \citep{krei2014}, and Spitzer \citep{frai2013} datasets.
    \label{fig:ts_modelfits_simple}}
\end{figure}

\begin{figure}
    \plotone{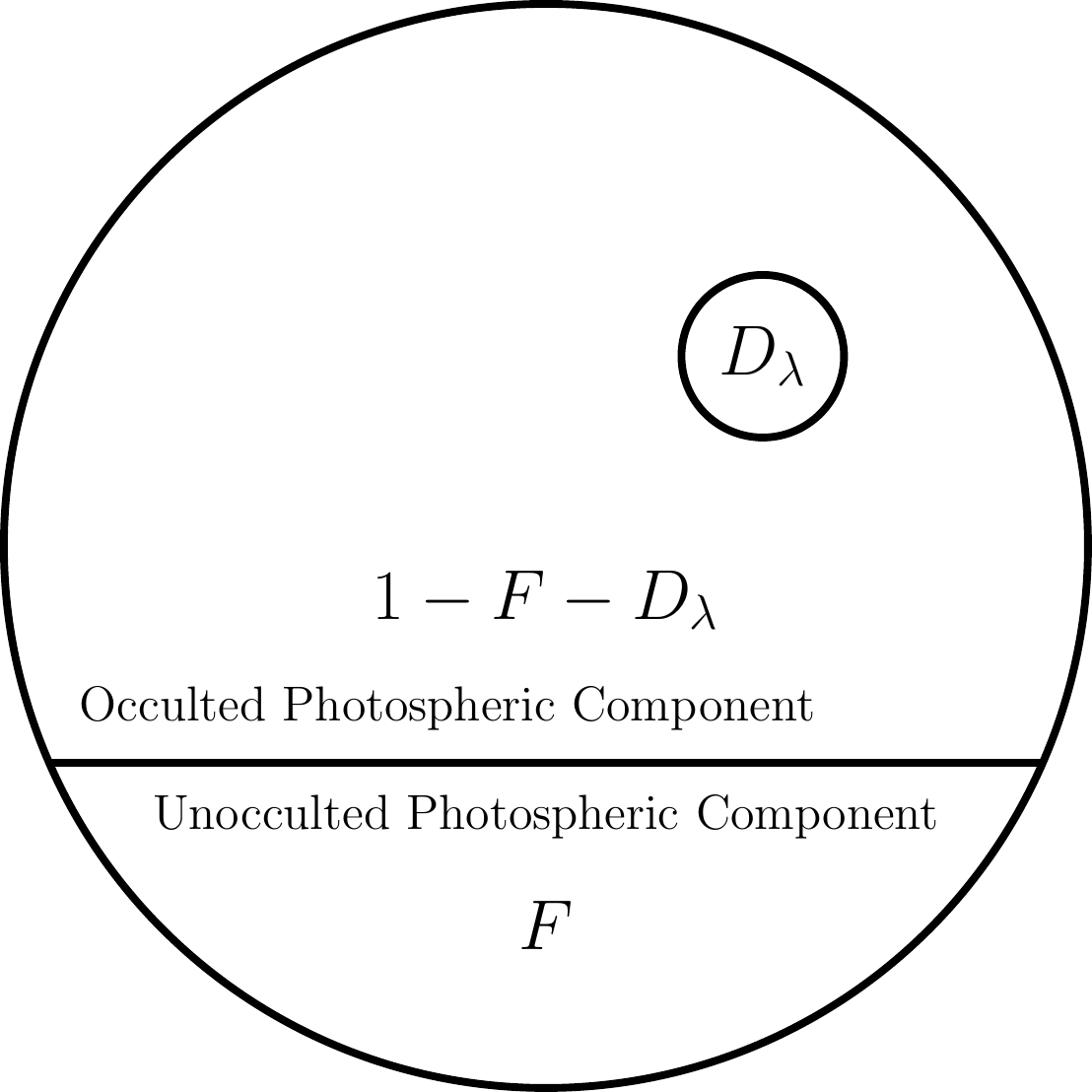}
    \caption{\textcolor{black}{ {\bf Schematic of composite photosphere and atmospheric transmission (CPAT) model.} In this model, the exoplanet blocks a wavelength-dependent fraction of the stellar disk $D_{\lambda}$. The transit chord probes a region with a characteristic stellar spectrum (the ``occulted'' spectrum), while a fraction of the stellar disk $F$, either continuous or not, is described by another (the ``unocculted'' spectrum). During the transit, the difference between the stellar spectra is imprinted in the observed transmission spectrum.}
    \label{fig:schematic}}
\end{figure}

\begin{figure}
   \epsscale{1}
    \plotone{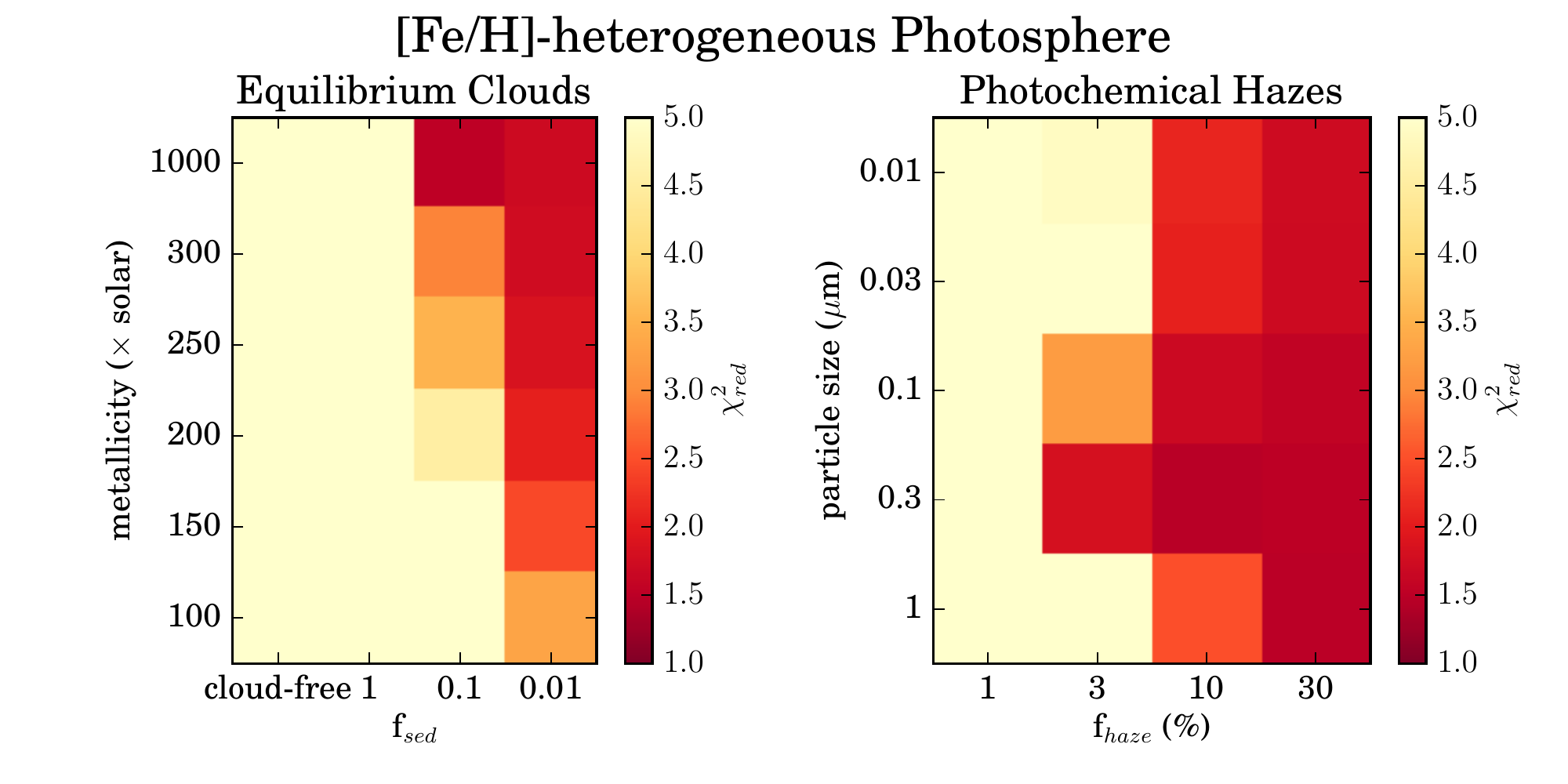}
    \caption{\textcolor{black}{ {\bf Reduced chi-squared maps for CPAT-absorber model fits.}  We find the best fits to the joint dataset for cloudy models with high metallicity ($1000\times$~solar) and thick clouds ($f_{sed}\sim0.01$--0.1) and hazy models with relatively large particles (0.1--1$~\micron$) and high haze-forming efficiencies (10--30\%). The free parameters in the fitting procedure are described in Section~\ref{subsec:CPAT-absorber} and their optimized values are provided in Tables~\ref{table:X2_CPAT-abs_cloud} and \ref{table:X2_CPAT-abs_haze} for the cloudy and hazy model grids \citep{morl2015}, respectively.}    
    \label{fig:X2_CPAT-abs}}
\end{figure}

\begin{figure}
   \epsscale{1}
    \plottwo{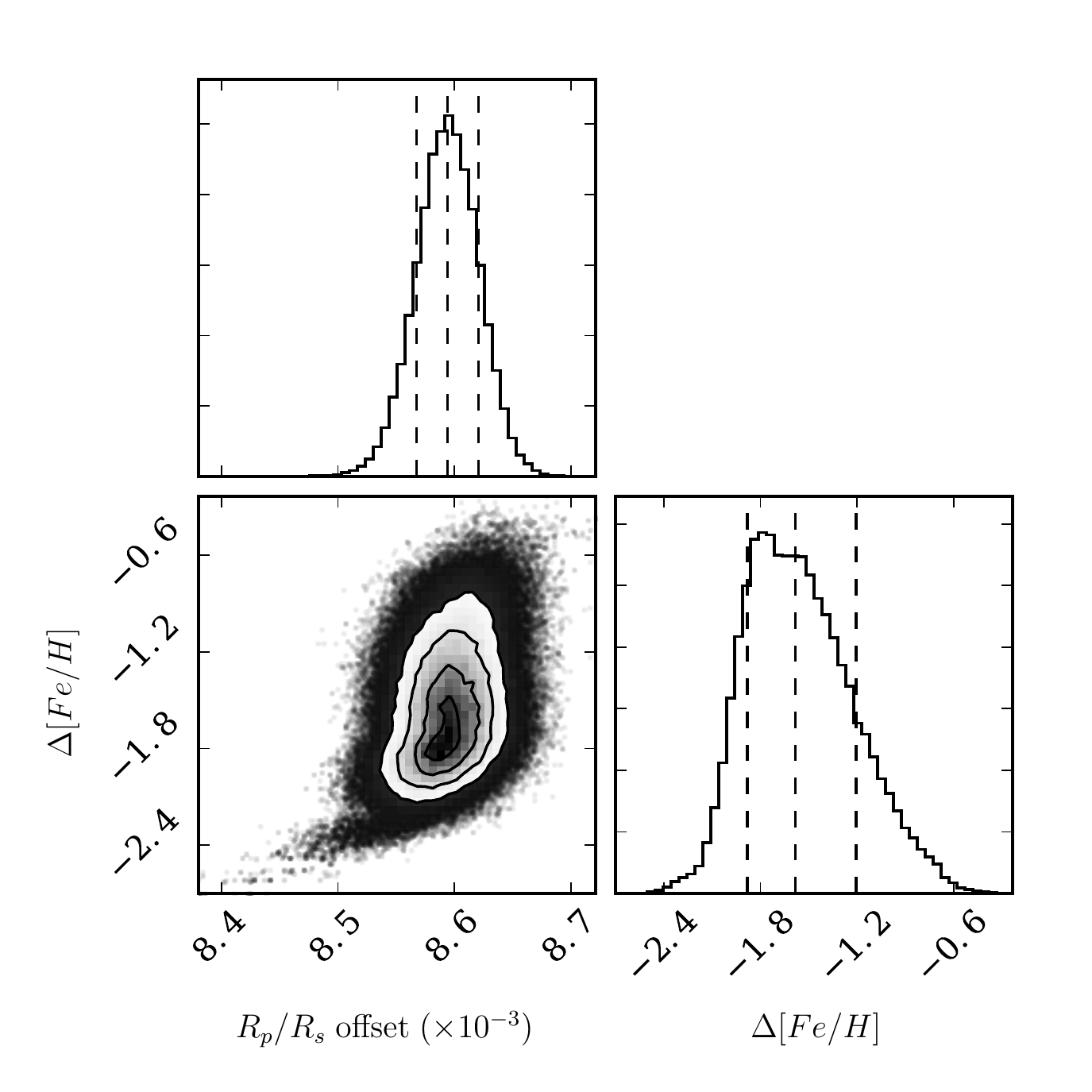}{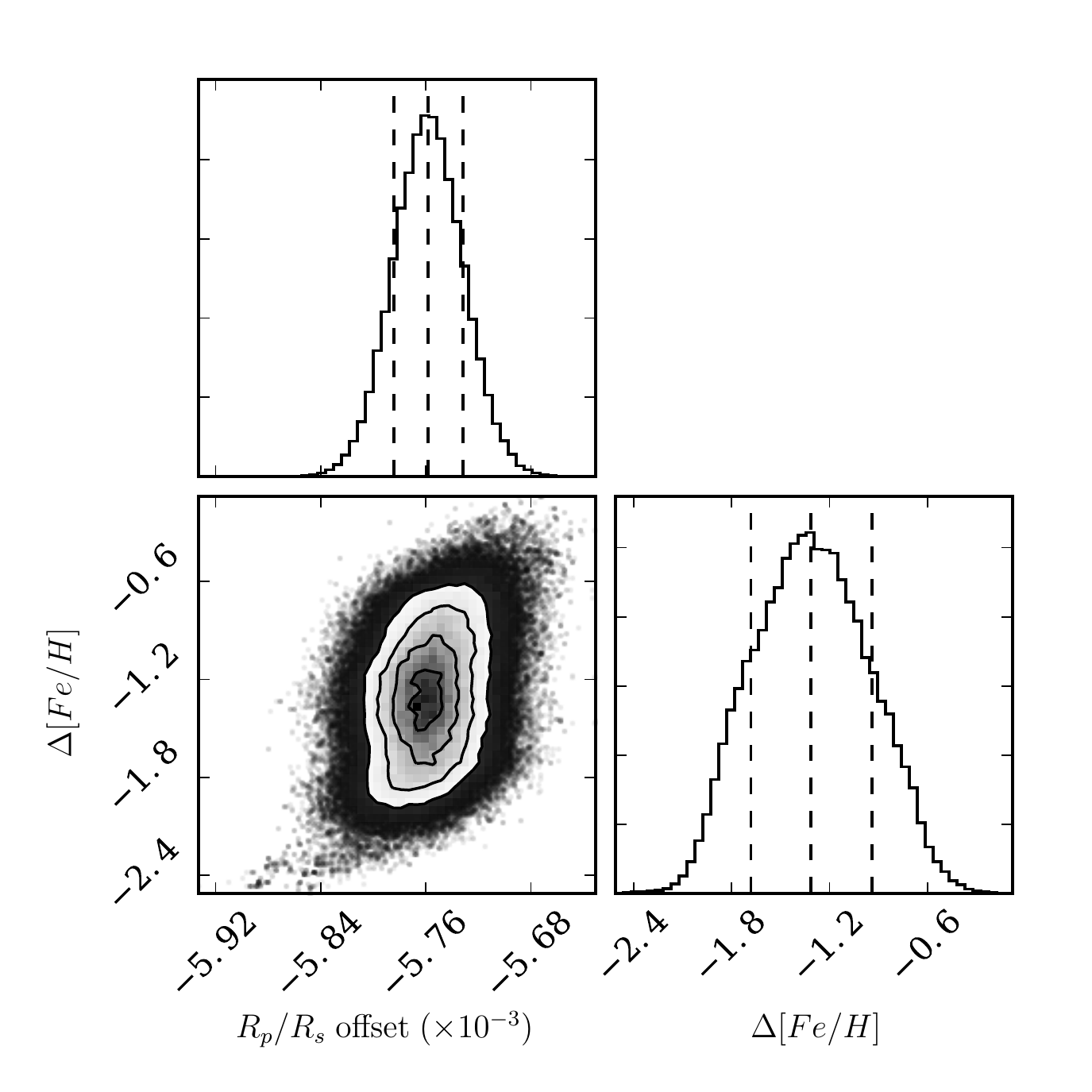}
    \caption{{\bf Posterior distributions for free parameters in CPAT-absorber models.}  The left panel illustrates the posterior distributions for the free parameters in the best-fitting cloud model ($1000\times$~solar, $f_{sed}=0.1$), and the right those for the best-fitting haze model ($r=0.3~\micron$, $f_{haze}=10\%$). Vertical dashed lines indicate the medians and 68\% confidence intervals.
    \label{fig:posteriors_CPAT-abs}}
\end{figure}

\begin{figure}
    \epsscale{0.9}
    \plotone{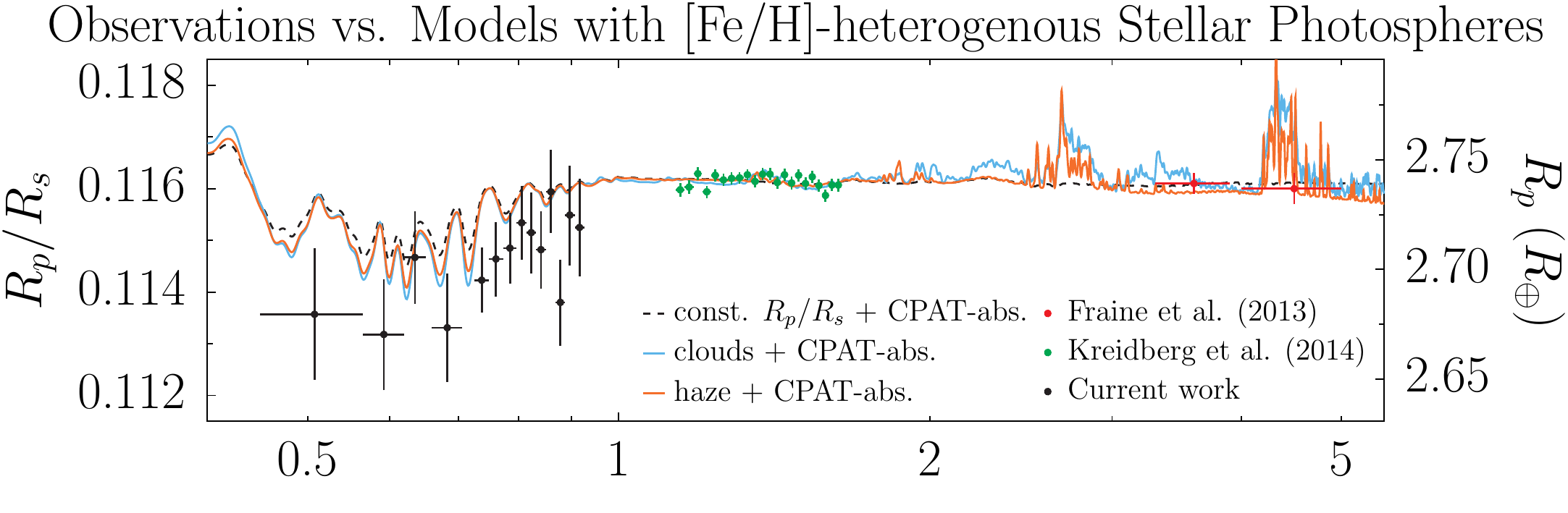}
    \plotone{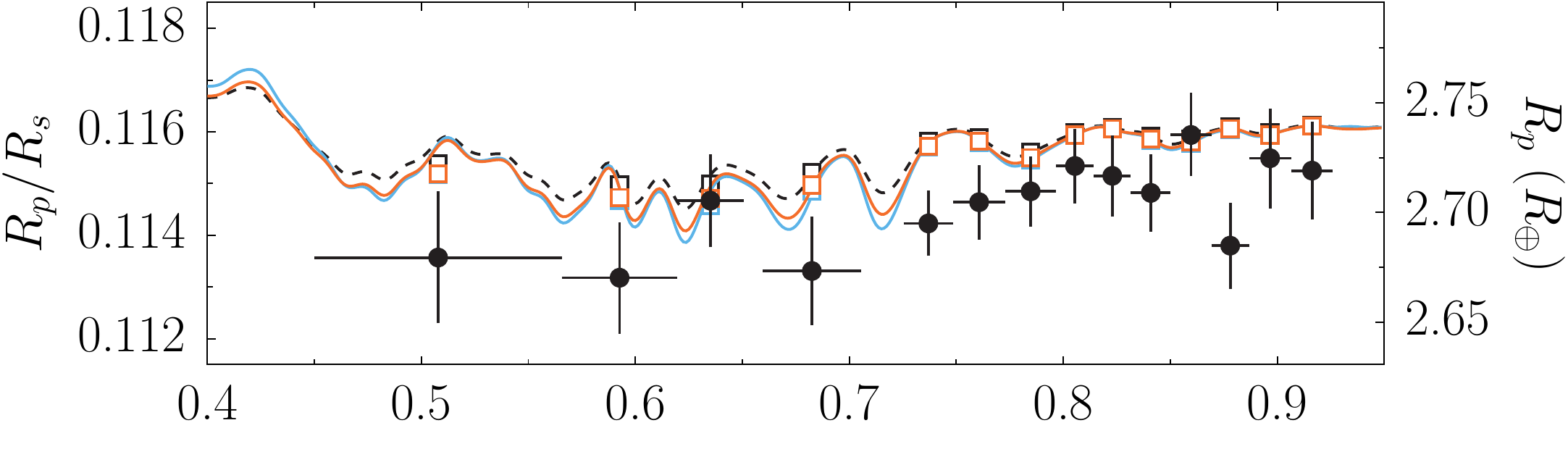}
    \plotone{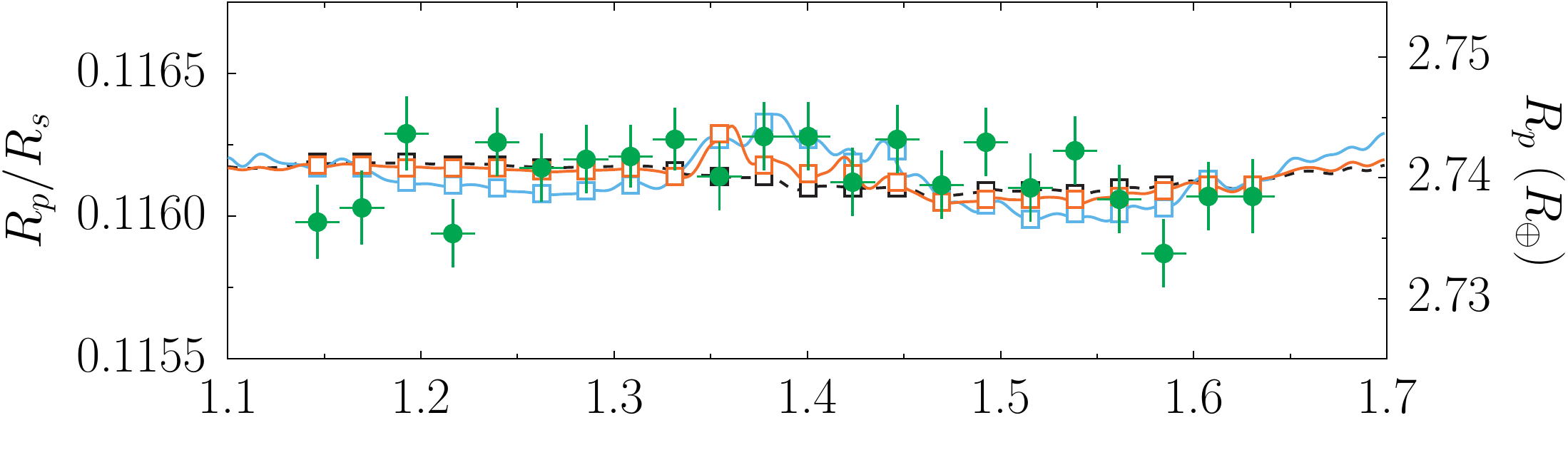}
    \plotone{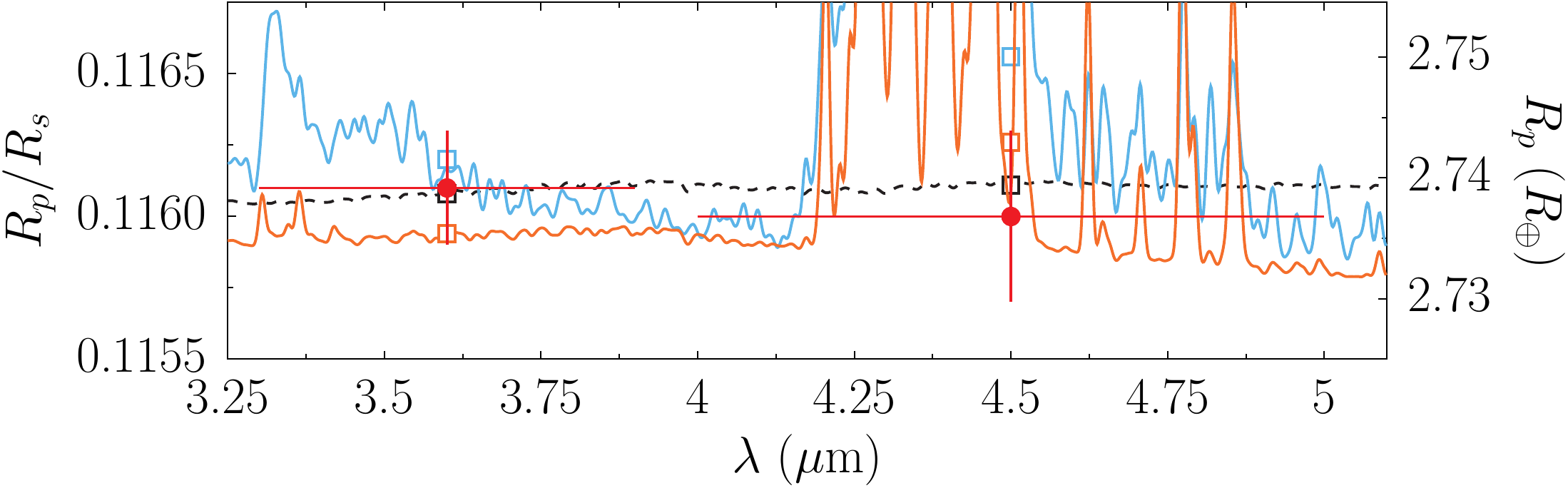}
    \caption{\textcolor{black}{ {\bf Best-fitting transmission spectra using CPAT-absorber model.} 
    The CPAT-absorber models reproduce both the optical and near-infrared measurements better than models for the exoplanetary atmosphere alone.
    The composite photospheres for the best-fitting models all include 3.2\% of the unocculted stellar disk that is described by a PHOENIX model with a lower metallicity, which imprints stellar absorption features on the observed transmission spectrum. The figure layout is the same as that of Figure~\ref{fig:ts_modelfits_simple}.}
     \label{fig:ts_modelfits_CPAT-abs}}
\end{figure}

\begin{figure}
   \epsscale{1}
    \plotone{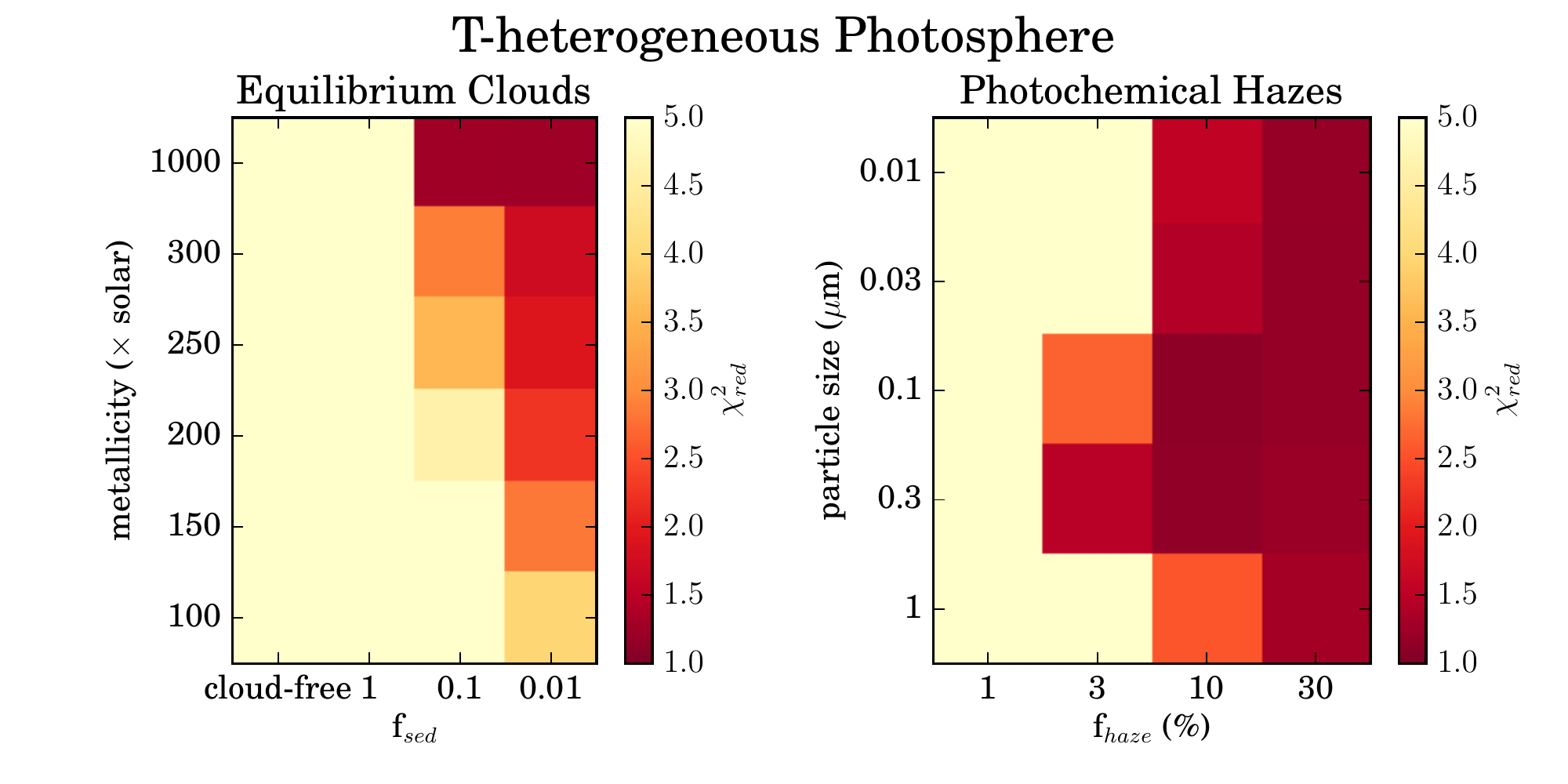}
    \caption{\textcolor{black}{ {\bf Reduced chi-squared maps for CPAT-temperature model fits.} As with the CPAT-absorber models, we find the best fits for cloudy models with high metallicity ($1000\times$~solar) and thick clouds ($f_{sed}\sim0.01$--0.1) and hazy models with relatively large particles (0.1--1~$\micron$) and high haze-forming efficiencies (10--30\%). The free parameters in the fitting procedure are described in Section~\ref{subsec:CPAT-temperature} and their optimized values are provided in Tables~\ref{table:X2_CPAT-temp_cloud} and \ref{table:X2_CPAT-temp_haze} for the cloudy and hazy grids \citep{morl2015}, respectively.}    
    \label{fig:X2_CPAT-temp}}
\end{figure}

\begin{figure}
   \epsscale{1}
    \plottwo{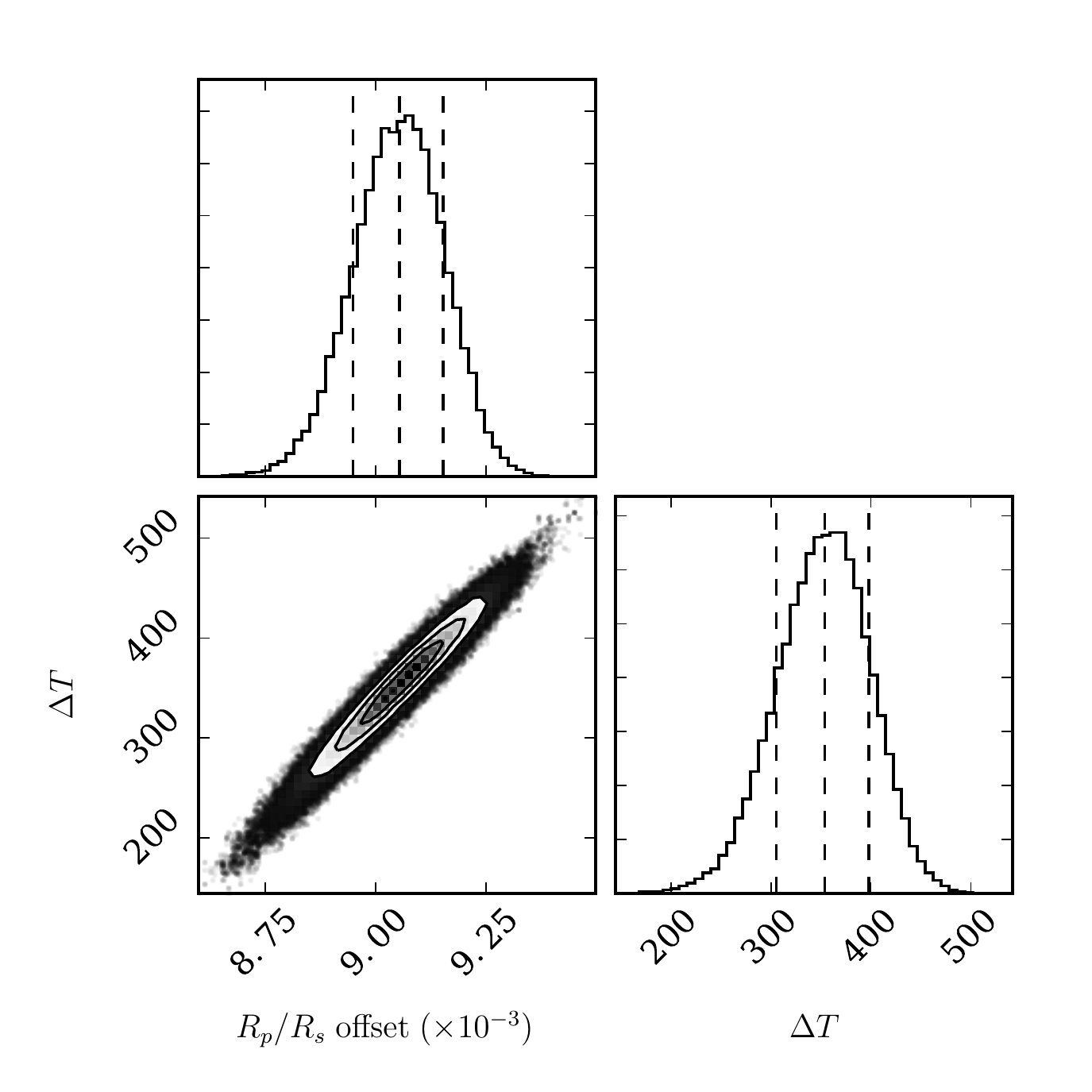}{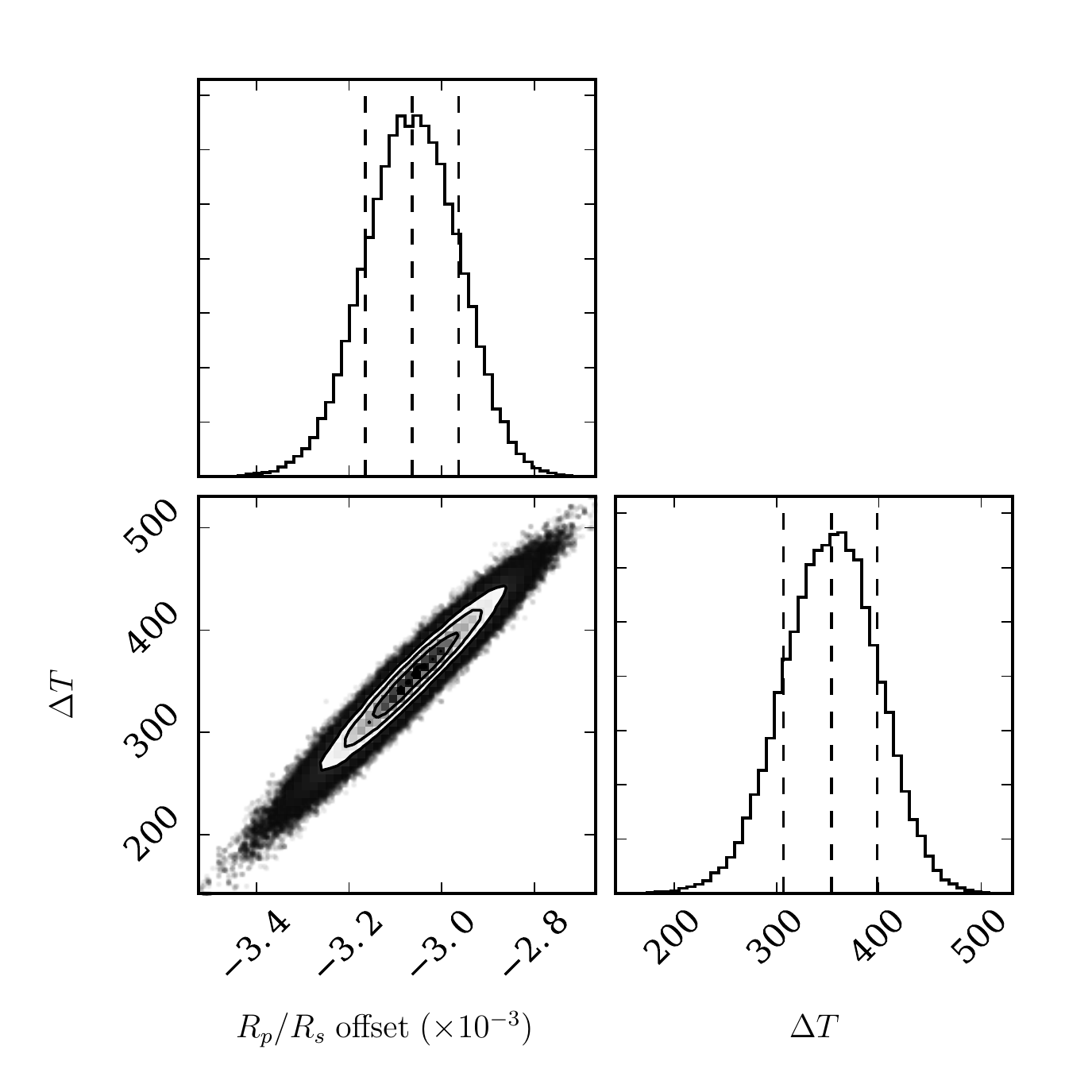}
    \caption{{\bf Posterior distributions for free parameters in CPAT-temperature models.}  The left panel illustrates the posterior distributions for the free parameters in the best-fitting cloud model ($1000\times$~solar, $f_{sed}=0.01$), and the right those for the best-fitting haze model ($r=0.1~\micron$, $f_{haze}=10\%$). Vertical dashed lines indicate the medians and 68\% confidence intervals.
    \label{fig:posteriors_CPAT-temp}}
\end{figure}

\begin{figure}
    \epsscale{0.9}
    \plotone{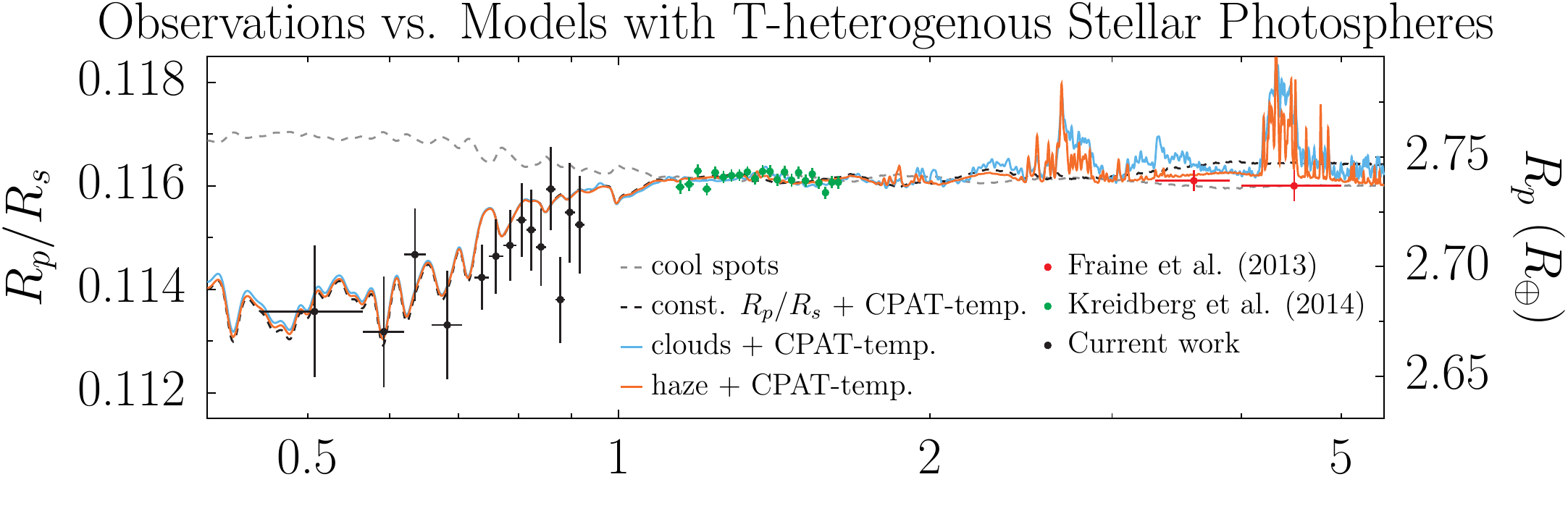}
    \plotone{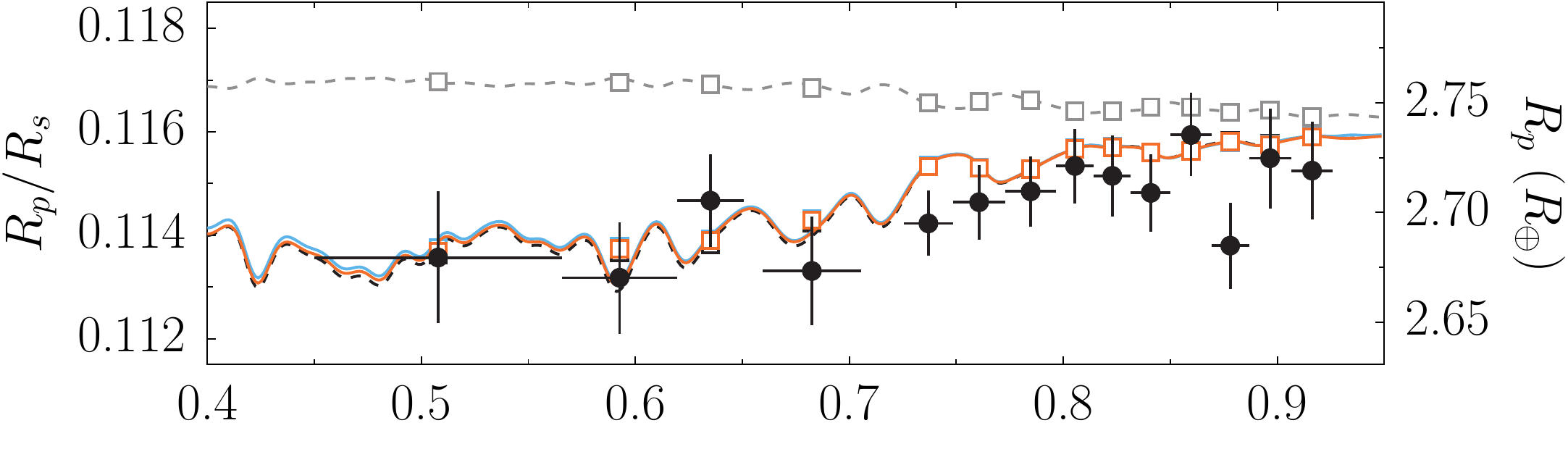}
    \plotone{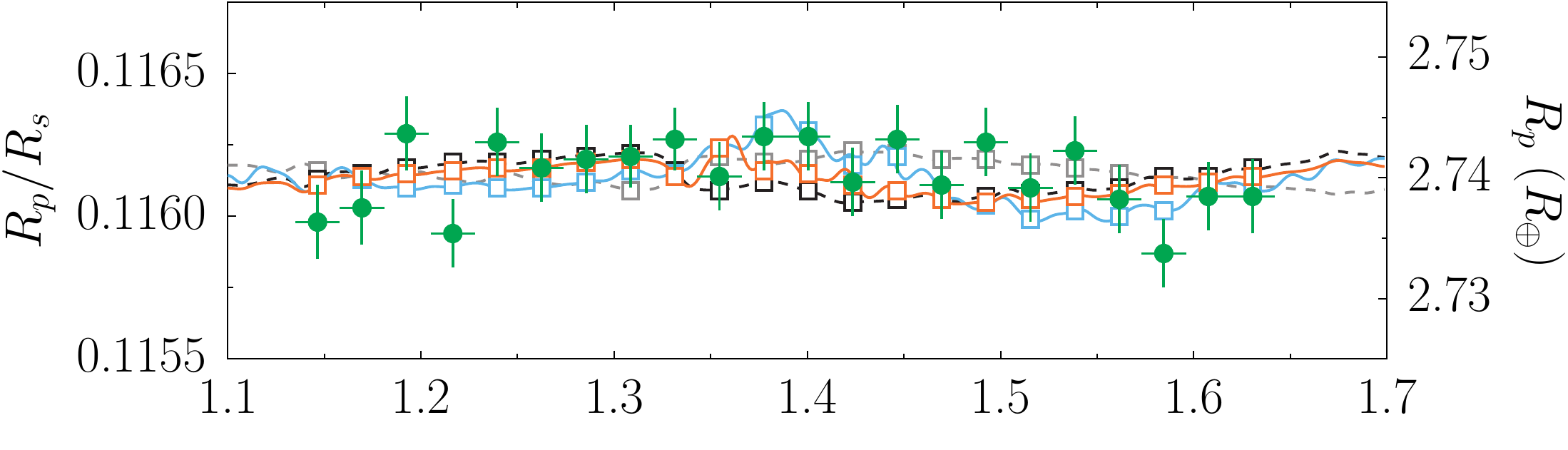}
    \plotone{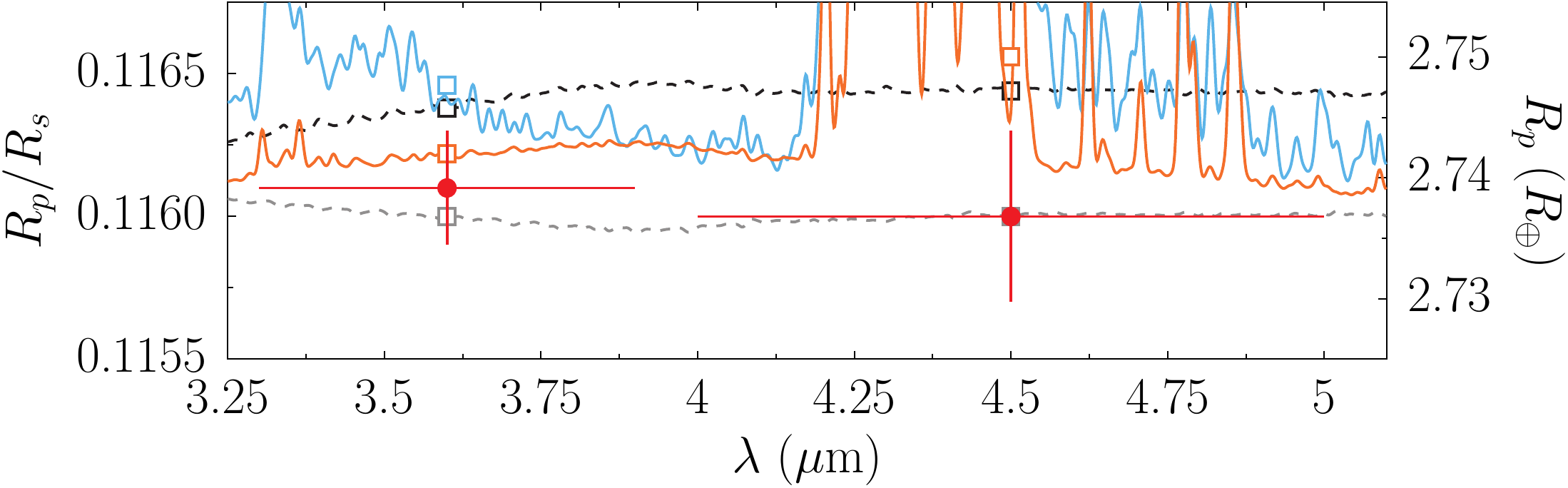}
    \caption{{\bf Best-fitting transmission spectra using CPAT-temperature model.}
    Models that include a temperature heterogeneity in the stellar photosphere provide the best fits to the near-infrared and optical measurements. 
    Bright regions covering $3.2\%$ of the unocculted stellar disk with a temperature contrast of $\sim350$~K can effectively decrease the flat $R_{p}/R_{s}$ (\textit{black}), cloud (\textit{blue}), and haze (\textit{orange}) models to the observed optical values, while only minimally altering the near-infrared spectrum.
    The effect of cool, unocculted starspots (\textit{gray dashed line}, see Section~\ref{subsec:CPAT-temperature}), however, does not match the optical data. 
     The figure layout is the same as that of Figure~\ref{fig:ts_modelfits_simple}.
     \label{fig:ts_modelfits_CPAT-temp}}
\end{figure}

\clearpage

\begin{deluxetable}{cccccccccc}
\rotate
\tablewidth{0pt}
\tabletypesize{\footnotesize}
\tablecaption{{\bf Observing log for GJ~1214b datasets from Magellan/IMACS}. \label{table:log}}
\tablehead{
\colhead{Transit} & \colhead{Date}   & \colhead{Obs.}         & \colhead{Camera} & \colhead{Disperser} & \colhead{Airmass}   & \colhead{Exposure} & \colhead{Readout +}     & \colhead{Frames}      & \colhead{Seeing} \\
\colhead{}            & \colhead{}          & \colhead{start / end}\tablenotemark{a} & \colhead{}             & \colhead{}                 & \colhead{}                & \colhead{times (s)}   & \colhead{overhead (s)}  & \colhead{}                  & \colhead{}
} 
\startdata
1                        & 2013 Apr 25          & 06:18--10:26           & f/4                          & 150 line/mm grat.   & 1.21--1.62               & 30--40                       & 34                                 & 235\tablenotemark{b}  & $\sim$0.6--0.8$''$\\ 
2                        & 2013 May 22        & 03:14--08:01           & f/4                          & 150 line/mm grat.   & 1.21--1.62               & 35--40                       & 31                                 & 243\tablenotemark{c}  & $\sim$0.7--0.9$''$\\ 
3                        & 2014 Apr 03         & 05:57--09:10           & f/2                          & 300 line/mm           & 1.21--1.88               & 63                              & 29                                  & 126                              & $\sim$0.5--1.0$''$\\  
                          &                             &                                  &                              & grism + 17.5          &                                  &                                   &                                       &                                   &                        \\
\enddata

\tablenotetext{a}{Dates and observation start / end times are given in UTC.}
\tablenotetext{b}{The last 14 frames, which were taken during twilight, displayed a systematic trend due to imperfect sky subtraction and were not included in the light curve analysis.}
\tablenotetext{c}{The last 40 frames displayed a systematic trend in the target's light curve and were not included in the light curve analysis.}
\end{deluxetable}

\begin{deluxetable}{ccccccc}
\tablewidth{0pt}
\tablecaption{{\bf Planet-to-star radius ratios}. \label{table:rprs}}
\tablehead{
\multicolumn{3}{c}{ Bin} & \multicolumn{4}{c}{$R_{p}/R_{s}$} \\
\colhead{N} & 
\colhead{$\lambda_{min}$} & 
\colhead{$\lambda_{max}$} & 
\colhead{Transit~1} & 
\colhead{Transit~2} & 
\colhead{Transit~3} & 
\colhead{Final}
}
\startdata
1 &  4500.0 &  5657.0 & 0.1121$^{+0.0020}_{-0.0020}$ & 0.1129$^{+0.0021}_{-0.0021}$ & 0.1172$^{+0.0026}_{-0.0027}$ & 0.1136$^{+0.0013}_{-0.0013}$ \\ 
2 &  5657.0 &  6196.0 & 0.1119$^{+0.0017}_{-0.0018}$ & 0.1139$^{+0.0017}_{-0.0018}$ & 0.1140$^{+0.0021}_{-0.0021}$ & 0.1132$^{+0.0011}_{-0.0011}$ \\ 
3 &  6196.0 &  6506.0 & 0.1153$^{+0.0015}_{-0.0015}$ & 0.1125$^{+0.0024}_{-0.0025}$ & 0.1149$^{+0.0013}_{-0.0013}$ & 0.1147$^{+0.0009}_{-0.0009}$ \\ 
4 &  6596.0 &  7054.0 & 0.1112$^{+0.0015}_{-0.0015}$ & 0.1153$^{+0.0024}_{-0.0025}$ & 0.1157$^{+0.0019}_{-0.0019}$ & 0.1133$^{+0.0011}_{-0.0010}$ \\ 
5 &  7254.0 &  7485.0 & 0.1147$^{+0.0012}_{-0.0012}$ & 0.1151$^{+0.0017}_{-0.0016}$ & 0.1138$^{+0.0008}_{-0.0008}$ & 0.1142$^{+0.0006}_{-0.0006}$ \\ 
6 &  7485.0 &  7728.0 & 0.1130$^{+0.0013}_{-0.0013}$ & 0.1161$^{+0.0014}_{-0.0013}$ & 0.1149$^{+0.0011}_{-0.0012}$ & 0.1146$^{+0.0007}_{-0.0007}$ \\ 
7 &  7728.0 &  7967.0 & 0.1155$^{+0.0013}_{-0.0013}$ & 0.1154$^{+0.0011}_{-0.0011}$ & 0.1138$^{+0.0012}_{-0.0011}$ & 0.1148$^{+0.0007}_{-0.0007}$ \\ 
8 &  7967.0 &  8142.0 & 0.1157$^{+0.0015}_{-0.0016}$ & 0.1174$^{+0.0013}_{-0.0012}$ & 0.1138$^{+0.0011}_{-0.0010}$ & 0.1153$^{+0.0007}_{-0.0007}$ \\ 
9 &  8142.0 &  8316.0 & 0.1150$^{+0.0018}_{-0.0019}$ & 0.1172$^{+0.0019}_{-0.0019}$ & 0.1146$^{+0.0010}_{-0.0010}$ & 0.1152$^{+0.0008}_{-0.0008}$ \\ 
10 &  8316.0 &  8502.0 & 0.1133$^{+0.0013}_{-0.0013}$ & 0.1162$^{+0.0016}_{-0.0016}$ & 0.1153$^{+0.0011}_{-0.0011}$ & 0.1148$^{+0.0007}_{-0.0008}$ \\ 
11 &  8502.0 &  8692.0 & 0.1150$^{+0.0014}_{-0.0014}$ & 0.1159$^{+0.0015}_{-0.0016}$ & 0.1168$^{+0.0013}_{-0.0014}$ & 0.1159$^{+0.0008}_{-0.0008}$ \\ 
12 &  8692.0 &  8868.0 & 0.1136$^{+0.0013}_{-0.0014}$ & 0.1161$^{+0.0018}_{-0.0018}$ & 0.1128$^{+0.0014}_{-0.0013}$ & 0.1138$^{+0.0008}_{-0.0008}$ \\ 
13 &  8868.0 &  9065.0 & 0.1164$^{+0.0017}_{-0.0017}$ & 0.1161$^{+0.0017}_{-0.0017}$ & 0.1141$^{+0.0016}_{-0.0017}$ & 0.1155$^{+0.0010}_{-0.0010}$ \\ 
14 &  9065.0 &  9260.0 & 0.1165$^{+0.0022}_{-0.0023}$ & 0.1153$^{+0.0016}_{-0.0016}$ & 0.1147$^{+0.0014}_{-0.0014}$ & 0.1153$^{+0.0009}_{-0.0010}$ \\ 
\enddata
\tablecomments{For individual transits, we report the medians and 68\% confidence intervals on the posterior distributions from the MCMC optimization procedure.
The final radius ratios are the weighted means of the three transit measurements (see Section \ref{subsec:combining_results} for details).}
\end{deluxetable}

\begin{deluxetable}{ccc}
\tablewidth{0pt}
\tablecaption{{\bf Mid-transit times from white light analyses}. \label{table:t0}}
\tablehead{
\colhead{Transit} & \colhead{Date (UTC)} & \colhead{t$_{0}$ (BJD$_{UTC}$)}
}
\startdata
1 & 2013 Apr 25 & 2456407.85423$^{+0.00008}_{-0.00008}$ \\ 
2 & 2013 May 22 & 2456434.72098$^{+0.00013}_{-0.00012}$ \\ 
3 & 2014 Apr 03 & 2456750.80196$^{+0.00008}_{-0.00008}$ \\ 
\enddata
\tablecomments{For the spectroscopic light curve analyses, the mid-transit times were fixed to these values. }
\end{deluxetable}

\clearpage

\begin{deluxetable}{ccccc}
\tabletypesize{\footnotesize}
\tablewidth{0pt}
\tablecaption{{\bf Model-fitting results for cloudy atmospheres considering only transmission through the exoplanet's limb.} \label{table:X2_simple_cloud}}
\tablehead{
\colhead{metallicity} & \colhead{$f_{sed}$\tablenotemark{a}} & \colhead{$\chi^{2}$} & \colhead{$DOF$} & \colhead{$\chi^{2}_{red}$} \\
\colhead{($\times$~solar)}                & \colhead{}                   & \colhead{}                 & \colhead{}            & \colhead{}
}
\startdata
100        & 0.01       & 215.3      & 37         & 5.82       \\ 
100        & 0.1        & 411.9      & 37         & 11.1       \\ 
100        & 1          & 3963       & 37         & 107        \\ 
100        & cloud-free & 6919       & 37         & 187        \\ 
150        & 0.01       & 168.0        & 37         & 4.54       \\ 
150        & 0.1        & 278.3      & 37         & 7.52       \\ 
150        & 1          & 2627       & 37         & 71.0         \\ 
150        & cloud-free & 4651       & 37         & 126        \\ 
200        & 0.01       & 142.7      & 37         & 3.86       \\ 
200        & 0.1        & 208.9      & 37         & 5.65       \\ 
200        & 1          & 1881       & 37         & 50.9      \\ 
200        & cloud-free & 3359       & 37         & 90.8       \\ 
250        & 0.01       & 127.1      & 37         & 3.44       \\ 
250        & 0.1        & 169.3      & 37         & 4.58       \\ 
250        & 1          & 1422       & 37         & 38.4       \\ 
250        & cloud-free & 2560       & 37         & 69.2       \\ 
300        & 0.01       & 116.7      & 37         & 3.16       \\ 
300        & 0.1        & 144.5      & 37         & 3.91       \\ 
300        & 1          & 1120       & 37         & 30.3       \\ 
300        & cloud-free & 2029       & 37         & 54.8       \\ 
1000       & 0.01       & 87.77      & 37         & 2.37       \\ 
1000       & 0.1        & 81.26      & 37         & 2.20        \\ 
1000       & 1          & 246.8      & 37         & 6.67       \\ 
1000       & cloud-free & 440.6      & 37         & 11.9       \\ 
\enddata

\tablenotetext{a}{Cloud-free models do not have an $f_{sed}$ value.}
\end{deluxetable}

\begin{deluxetable}{ccccc}
\tabletypesize{\footnotesize}
\tablewidth{0pt}
\tablecaption{{\bf Model-fitting results for hazy atmospheres considering only transmission through the exoplanet's limb.} \label{table:X2_simple_haze}}
\tablehead{
\colhead{$r$}               & \colhead{$f_{haze}$} & \colhead{$\chi^{2}$} & \colhead{$DOF$} & \colhead{$\chi^{2}_{red}$} \\
\colhead{$(\micron)$}  & \colhead{(\%)}           & \colhead{}                 & \colhead{}            & \colhead{}
}
\startdata
0.01       & 1          & 916.1      & 37         & 24.8       \\ 
0.01       & 3          & 394.4      & 37         & 10.7       \\ 
0.01       & 10         & 115.1      & 37         & 3.11       \\ 
0.01       & 30         & 82.89      & 37         & 2.24       \\ 
0.03       & 1          & 753.6      & 37         & 20.4       \\ 
0.03       & 3          & 376.6      & 37         & 10.2       \\ 
0.03       & 10         & 106.7      & 37         & 2.88       \\ 
0.03       & 30         & 81.84      & 37         & 2.21       \\ 
0.1        & 1          & 338.9      & 37         & 9.16       \\ 
0.1        & 3          & 175.0        & 37         & 4.73       \\ 
0.1        & 10         & 82.49      & 37         & 2.23       \\ 
0.1        & 30         & 77.37      & 37         & 2.09       \\ 
0.3        & 1          & 1119       & 37         & 30.2       \\ 
0.3        & 3          & 90.66      & 37         & 2.45       \\ 
0.3        & 10         & 74.13      & 37         & 2.00        \\ 
0.3        & 30         & 75.58      & 37         & 2.04       \\ 
1          & 1          & 5578       & 37         & 151        \\ 
1          & 3          & 1734       & 37         & 46.9       \\ 
1          & 10         & 117.0        & 37         & 3.16       \\ 
1          & 30         & 75.77      & 37         & 2.05       \\ 
\enddata
\end{deluxetable}


\begin{deluxetable}{ccccccc}
\tabletypesize{\footnotesize}
\tablewidth{0pt}
\tablecaption{{\bf Results from MCMC fits to CPAT-absorber models for cloudy atmospheres}. \label{table:X2_CPAT-abs_cloud}}
\tablehead{
\colhead{metallicity} & \colhead{$f_{sed}$\tablenotemark{a}} & \colhead{$\Delta[Fe/H]$} & \colhead{$(R_{p}/R_{s})_{o}$} & \colhead{$\chi^{2}$} & \colhead{$DOF$} & \colhead{$\chi^{2}_{red}$} \\
\colhead{($\times$~solar)}                & \colhead{}                  & \colhead{}                      & \colhead{(ppm)}                             & \colhead{}                 & \colhead{}            & \colhead{}
}
\startdata
100        & 0.01       & -3.38$^{+0.55}_{-0.29}$ & -1364$^{+155}_{-96}$ & 119.7      & 36         & 3.32       \\ 
100        & 0.1        & -2.23$^{+0.14}_{-0.15}$ & -32$^{+42}_{-35}$ & 354.6      & 36         & 9.85       \\ 
100        & 1          & -1.72$^{+0.12}_{-0.14}$ & 1956$^{+27}_{-25}$ & 3914       & 36         & 109        \\ 
100        & cloud-free & -1.53$^{+0.11}_{-0.18}$ & 2392$^{+26}_{-25}$ & 6863       & 36         & 191        \\ 
150        & 0.01       & -2.80$^{+0.18}_{-0.15}$ & 601$^{+64}_{-55}$ & 87.68      & 36         & 2.44       \\ 
150        & 0.1        & -2.19$^{+0.16}_{-0.17}$ & 1638$^{+42}_{-33}$ & 227.9      & 36         & 6.33       \\ 
150        & 1          & -1.68$^{+0.13}_{-0.14}$ & 3288$^{+25}_{-26}$ & 2580       & 36         & 71.7       \\ 
150        & cloud-free & -1.46$^{+0.13}_{-0.23}$ & 3646$^{+25}_{-25}$ & 4590       & 36         & 127        \\ 
200        & 0.01       & -2.71$^{+0.15}_{-0.14}$ & 1990$^{+51}_{-54}$ & 73.65      & 36         & 2.05       \\ 
200        & 0.1        & -2.14$^{+0.18}_{-0.19}$ & 2900$^{+40}_{-31}$ & 163.5      & 36         & 4.54       \\ 
200        & 1          & -1.65$^{+0.14}_{-0.16}$ & 4340$^{+26}_{-25}$ & 1834       & 36         & 50.9       \\ 
200        & cloud-free & -1.42$^{+0.21}_{-0.16}$ & 4645$^{+25}_{-25}$ & 3306       & 36         & 91.8       \\ 
250        & 0.01       & -2.65$^{+0.15}_{-0.15}$ & 3075$^{+52}_{-52}$ & 66.64      & 36         & 1.85       \\ 
250        & 0.1        & -2.09$^{+0.19}_{-0.21}$ & 3895$^{+37}_{-31}$ & 127.1      & 36         & 3.53       \\ 
250        & 1          & -1.63$^{+0.15}_{-0.16}$ & 5180$^{+27}_{-24}$ & 1378       & 36         & 38.3       \\ 
250        & cloud-free & -1.41$^{+0.21}_{-0.16}$ & 5450$^{+26}_{-25}$ & 2508       & 36         & 69.7       \\ 
300        & 0.01       & -2.60$^{+0.15}_{-0.15}$ & 3940$^{+51}_{-52}$ & 62.5       & 36         & 1.74       \\ 
300        & 0.1        & -2.03$^{+0.20}_{-0.23}$ & 4688$^{+36}_{-30}$ & 105.6      & 36         & 2.93       \\ 
300        & 1          & -1.62$^{+0.15}_{-0.18}$ & 5854$^{+25}_{-26}$ & 1074       & 36         & 29.8       \\ 
300        & cloud-free & -1.41$^{+0.21}_{-0.17}$ & 6098$^{+26}_{-24}$ & 1981       & 36         & 55         \\ 
1000       & 0.01       & -1.98$^{+0.23}_{-0.54}$ & 8231$^{+54}_{-37}$ & 61.43      & 36         & 1.71       \\ 
1000       & 0.1        & -1.58$^{+0.28}_{-0.37}$ & 8594$^{+28}_{-26}$ & 54.67      & 36         & 1.52       \\ 
1000       & 1          & -1.54$^{+0.20}_{-0.24}$ & 9205$^{+26}_{-24}$ & 212.9      & 36         & 5.91       \\ 
1000       & cloud-free & -1.39$^{+0.21}_{-0.22}$ & 9339$^{+25}_{-26}$ & 404.9      & 36         & 11.2       \\ 
\enddata

\tablenotetext{a}{Cloud-free models do not have an $f_{sed}$ value.}
\tablecomments{The free parameters in the fitting procedure, $\Delta[Fe/H]$ and $(R_{p}/R_{s})_{o}$, are described in Section~\ref{subsec:CPAT-absorber}. We report the median and 68\% confidence intervals from the MCMC optimization procedure for each free parameter. With 38 data points and 2 fitted parameters, we assume 36 DOF when calculating the $\chi^{2}_{red}$ statistic.}
\end{deluxetable}

\begin{deluxetable}{ccccccc}
\tabletypesize{\footnotesize}
\tablewidth{0pt}
\tablecaption{{\bf Results from MCMC fits to CPAT-absorber models for hazy atmospheres}. \label{table:X2_CPAT-abs_haze}}
\tablehead{
\colhead{$r$}               & \colhead{$f_{haze}$} & \colhead{$\Delta[Fe/H]$} & \colhead{$(R_{p}/R_{s})_{o}$} & \colhead{$\chi^{2}$} & \colhead{$DOF$} & \colhead{$\chi^{2}_{red}$} \\
\colhead{$(\micron)$}  & \colhead{(\%)}           & \colhead{}                      & \colhead{(ppm)}                             & \colhead{}                 & \colhead{}            & \colhead{}
}
\startdata
0.01       & 1          & -4.16$^{+0.02}_{-0.04}$ & -4007$^{+28}_{-28}$ & 512.4      & 36         & 14.2       \\ 
0.01       & 3          & -3.93$^{+0.15}_{-0.17}$ & -4626$^{+49}_{-54}$ & 175.6      & 36         & 4.88       \\ 
0.01       & 10         & -2.57$^{+0.15}_{-0.15}$ & -4064$^{+52}_{-54}$ & 76.34      & 36         & 2.12       \\ 
0.01       & 30         & -1.65$^{+0.41}_{-0.43}$ & -3201$^{+31}_{-28}$ & 61.98      & 36         & 1.72       \\ 
0.03       & 1          & -4.16$^{+0.03}_{-0.04}$ & -4777$^{+27}_{-31}$ & 333.8      & 36         & 9.27       \\ 
0.03       & 3          & -3.67$^{+0.18}_{-0.17}$ & -4743$^{+53}_{-56}$ & 215        & 36         & 5.97       \\ 
0.03       & 10         & -2.47$^{+0.15}_{-0.18}$ & -3879$^{+54}_{-60}$ & 74.38      & 36         & 2.07       \\ 
0.03       & 30         & -1.61$^{+0.40}_{-0.44}$ & -2928$^{+31}_{-27}$ & 61.51      & 36         & 1.71       \\ 
0.1        & 1          & -3.55$^{+0.18}_{-0.18}$ & -5320$^{+55}_{-57}$ & 184.2      & 36         & 5.12       \\ 
0.1        & 3          & -2.77$^{+0.15}_{-0.15}$ & -5538$^{+53}_{-52}$ & 115.4      & 36         & 3.21       \\ 
0.1        & 10         & -1.77$^{+0.45}_{-0.35}$ & -3871$^{+35}_{-28}$ & 61.18      & 36         & 1.7        \\ 
0.1        & 30         & -1.35$^{+0.35}_{-0.44}$ & -2965$^{+28}_{-26}$ & 56.68      & 36         & 1.57       \\ 
0.3        & 1          & -1.69$^{+0.15}_{-0.17}$ & -3184$^{+26}_{-26}$ & 1079       & 36         & 30         \\ 
0.3        & 3          & -1.76$^{+0.26}_{-0.29}$ & -6049$^{+29}_{-27}$ & 65.6       & 36         & 1.82       \\ 
0.3        & 10         & -1.31$^{+0.35}_{-0.39}$ & -5758$^{+27}_{-26}$ & 53.34      & 36         & 1.48       \\ 
0.3        & 30         & -1.24$^{+0.36}_{-0.33}$ & -4403$^{+26}_{-26}$ & 54.21      & 36         & 1.51       \\ 
1          & 1          & -1.61$^{+0.10}_{-0.12}$ & -635$^{+25}_{-25}$ & 5518       & 36         & 153        \\ 
1          & 3          & -1.54$^{+0.15}_{-0.19}$ & -2976$^{+26}_{-25}$ & 1691       & 36         & 47         \\ 
1          & 10         & -1.29$^{+0.26}_{-0.28}$ & -6518$^{+26}_{-25}$ & 90.08      & 36         & 2.5        \\ 
1          & 30         & -1.21$^{+0.32}_{-0.31}$ & -7258$^{+26}_{-26}$ & 53.73      & 36         & 1.49       \\ 
\enddata
\tablecomments{
The free parameters in the fitting procedure, $\Delta[Fe/H]$ and $(R_{p}/R_{s})_{o}$, are described in Section~\ref{subsec:CPAT-absorber}. We report the median and 68\% confidence intervals from the MCMC optimization procedure for each free parameter. With 38 data points and 2 fitted parameters, we assume 36 DOF when calculating the $\chi^{2}_{red}$ statistic.}
\end{deluxetable}

\clearpage

\begin{deluxetable}{ccccccccc}
\tabletypesize{\footnotesize}
\tablewidth{0pt}
\tablecaption{{\bf Results from MCMC fits to CPAT-temperature models for cloudy atmospheres}. \label{table:X2_CPAT-temp_cloud}}
\tablehead{
\colhead{metallicity} & \colhead{$f_{sed}$\tablenotemark{a}} & \colhead{$\Delta T$} & \colhead{$(R_{p}/R_{s})_{o}$} & \colhead{$\chi^{2}$} & \colhead{$DOF$} & \colhead{$\chi^{2}_{red}$} \\
\colhead{($\times$~solar)}                  & \colhead{}                 & \colhead{(K)}          & \colhead{(ppm)}                             & \colhead{}                 & \colhead{}            & \colhead{}
}
\startdata
100        & 0.01       & 417$^{+39}_{-41}$ & -45$^{+91}_{-89}$ & 142.9      & 36         & 3.97       \\ 
100        & 0.1        & 328$^{+43}_{-43}$ & 759$^{+99}_{-90}$ & 368.9      & 36         & 10.2       \\ 
100        & 1          & 344$^{+30}_{-32}$ & 2735$^{+73}_{-68}$ & 3887       & 36         & 108        \\ 
100        & cloud-free & 389$^{+29}_{-40}$ & 3266$^{+75}_{-83}$ & 6785       & 36         & 188        \\ 
150        & 0.01       & 408$^{+43}_{-41}$ & 1734$^{+98}_{-90}$ & 102        & 36         & 2.83       \\ 
150        & 0.1        & 329$^{+42}_{-45}$ & 2424$^{+95}_{-96}$ & 235.7      & 36         & 6.55       \\ 
150        & 1          & 331$^{+34}_{-34}$ & 4039$^{+82}_{-71}$ & 2563       & 36         & 71.2       \\ 
150        & cloud-free & 359$^{+30}_{-28}$ & 4457$^{+69}_{-66}$ & 4555       & 36         & 127        \\ 
200        & 0.01       & 399$^{+43}_{-41}$ & 3073$^{+97}_{-92}$ & 80.93      & 36         & 2.25       \\ 
200        & 0.1        & 333$^{+43}_{-45}$ & 3689$^{+95}_{-97}$ & 166.3      & 36         & 4.62       \\ 
200        & 1          & 324$^{+39}_{-35}$ & 5076$^{+84}_{-81}$ & 1827       & 36         & 50.8       \\ 
200        & cloud-free & 347$^{+29}_{-30}$ & 5426$^{+69}_{-66}$ & 3274       & 36         & 90.9       \\ 
250        & 0.01       & 392$^{+44}_{-42}$ & 4126$^{+95}_{-98}$ & 69.26      & 36         & 1.92       \\ 
250        & 0.1        & 332$^{+44}_{-44}$ & 4676$^{+95}_{-98}$ & 128.5      & 36         & 3.57       \\ 
250        & 1          & 317$^{+40}_{-38}$ & 5903$^{+87}_{-84}$ & 1368       & 36         & 38         \\ 
250        & cloud-free & 338$^{+32}_{-32}$ & 6213$^{+72}_{-72}$ & 2489       & 36         & 69.1       \\ 
300        & 0.01       & 384$^{+45}_{-42}$ & 4957$^{+102}_{-93}$ & 61.85      & 36         & 1.72       \\ 
300        & 0.1        & 332$^{+46}_{-42}$ & 5465$^{+94}_{-99}$ & 104.1      & 36         & 2.89       \\ 
300        & 1          & 315$^{+41}_{-39}$ & 6572$^{+91}_{-87}$ & 1070       & 36         & 29.7       \\ 
300        & cloud-free & 329$^{+35}_{-33}$ & 6842$^{+78}_{-76}$ & 1967       & 36         & 54.6       \\ 
1000       & 0.01       & 354$^{+46}_{-47}$ & 9054$^{+110}_{-94}$ & 45.15      & 36         & 1.25       \\ 
1000       & 0.1        & 324$^{+47}_{-47}$ & 9331$^{+103}_{-103}$ & 46.13      & 36         & 1.28       \\ 
1000       & 1          & 309$^{+47}_{-42}$ & 9910$^{+99}_{-96}$ & 210.9      & 36         & 5.86       \\ 
1000       & cloud-free & 298$^{+47}_{-42}$ & 10019$^{+100}_{-92}$ & 404.8      & 36         & 11.2       \\ 
\enddata

\tablenotetext{a}{Cloud-free models do not have an $f_{sed}$ value.}
\tablecomments{
The free parameters in the fitting procedure, $\Delta T$ and $(R_{p}/R_{s})_{o}$, are described in Section~\ref{subsec:CPAT-temperature}. We report the median and 68\% confidence intervals from the MCMC optimization procedure for each free parameter. With 38 data points and 2 fitted parameters, we assume 36 DOF when calculating the $\chi^{2}_{red}$ statistic.}
\end{deluxetable}

\begin{deluxetable}{ccccccccc}
\tabletypesize{\footnotesize}
\tablewidth{0pt}
\tablecaption{{\bf Results from MCMC fits to CPAT-temperature models for hazy atmospheres}. \label{table:X2_CPAT-temp_haze}}
\tablehead{
\colhead{$r$}               & \colhead{$f_{haze}$} & \colhead{$\Delta T$} & \colhead{$(R_{p}/R_{s})_{o}$} & \colhead{$\chi^{2}$} & \colhead{$DOF$} & \colhead{$\chi^{2}_{red}$} \\
\colhead{$(\micron)$}  & \colhead{(\%)}            & \colhead{(K)}         & \colhead{(ppm)}                             & \colhead{}                 & \colhead{}            & \colhead{}
}
\startdata
0.01       & 1          & 630$^{+31}_{-29}$ & -1974$^{+82}_{-73}$ & 673        & 36         & 18.7       \\ 
0.01       & 3          & 546$^{+36}_{-36}$ & -2860$^{+88}_{-89}$ & 255        & 36         & 7.08       \\ 
0.01       & 10         & 404$^{+46}_{-38}$ & -3014$^{+98}_{-91}$ & 55.96      & 36         & 1.55       \\ 
0.01       & 30         & 345$^{+48}_{-45}$ & -2416$^{+106}_{-98}$ & 42.93      & 36         & 1.19       \\ 
0.03       & 1          & 616$^{+34}_{-29}$ & -2780$^{+83}_{-79}$ & 542.8      & 36         & 15.1       \\ 
0.03       & 3          & 527$^{+39}_{-37}$ & -3089$^{+87}_{-93}$ & 252.2      & 36         & 7.01       \\ 
0.03       & 10         & 392$^{+46}_{-41}$ & -2883$^{+108}_{-89}$ & 51.3       & 36         & 1.43       \\ 
0.03       & 30         & 344$^{+46}_{-47}$ & -2145$^{+103}_{-103}$ & 42.46      & 36         & 1.18       \\ 
0.1        & 1          & 487$^{+38}_{-38}$ & -3791$^{+91}_{-87}$ & 227.4      & 36         & 6.32       \\ 
0.1        & 3          & 445$^{+40}_{-39}$ & -4334$^{+90}_{-91}$ & 96.09      & 36         & 2.67       \\ 
0.1        & 10         & 354$^{+46}_{-46}$ & -3064$^{+101}_{-102}$ & 40.52      & 36         & 1.13       \\ 
0.1        & 30         & 329$^{+51}_{-45}$ & -2222$^{+107}_{-101}$ & 42.54      & 36         & 1.18       \\ 
0.3        & 1          & 318$^{+41}_{-37}$ & -2459$^{+89}_{-82}$ & 1071       & 36         & 29.7       \\ 
0.3        & 3          & 325$^{+47}_{-44}$ & -5304$^{+97}_{-102}$ & 53.34      & 36         & 1.48       \\ 
0.3        & 10         & 322$^{+48}_{-48}$ & -5029$^{+107}_{-102}$ & 41.24      & 36         & 1.15       \\ 
0.3        & 30         & 314$^{+52}_{-45}$ & -3691$^{+105}_{-104}$ & 43.98      & 36         & 1.22       \\ 
1          & 1          & 358$^{+27}_{-28}$ & 174$^{+62}_{-68}$ & 5470       & 36         & 152        \\ 
1          & 3          & 297$^{+44}_{-39}$ & -2298$^{+93}_{-87}$ & 1689       & 36         & 46.9       \\ 
1          & 10         & 272$^{+51}_{-51}$ & -5893$^{+107}_{-109}$ & 91.88      & 36         & 2.55       \\ 
1          & 30         & 302$^{+51}_{-47}$ & -6572$^{+110}_{-102}$ & 47.19      & 36         & 1.31       \\ 
\enddata
\tablecomments{
The free parameters in the fitting procedure, $\Delta T$ and $(R_{p}/R_{s})_{o}$, are described in Section~\ref{subsec:CPAT-temperature}. We report the median and 68\% confidence intervals from the MCMC optimization procedure for each free parameter. With 38 data points and 2 fitted parameters, we assume 36 DOF when calculating the $\chi^{2}_{red}$ statistic.}
\end{deluxetable}

\clearpage



{\it Facilities:} \facility{Magellan:Baade (IMACS)}




\end{document}